\documentclass[a4paper,11pt]{article}
\pdfoutput=1 

\usepackage{jheppub} 

\usepackage[T1]{fontenc} 

\usepackage{caption,subcaption}
\usepackage{tikz}
\usetikzlibrary{calc}
\usetikzlibrary{arrows,decorations.pathmorphing,patterns}
\usepackage{soul}

\numberwithin{equation}{section}
\usepackage{version}
\usepackage{mathrsfs} 

\title{\boldmath Renormalization of spin-one asymptotic charges in AdS$_D$}

\author[a,1]{Andrea Campoleoni,\note{Research associate of the Fund for Scientific Research – FNRS, Belgium.}}
\author[a,2]{Arnaud Delfante,\note{FRIA grantee of the Fund for Scientific Research – FNRS, Belgium.}}
\author[b]{Dario Francia}
\author[c,d]{and Carlo Heissenberg}

\affiliation[a]{Service de Physique de l'Univers, Champs et Gravitation,\\
Universit{\'e} de Mons -- UMONS,
20 place du Parc, 7000 Mons, Belgium}
\affiliation[b]{Roma Tre University and INFN Roma Tre\\ 
via della Vasca Navale, 84 I-00146 Roma, Italy}
\affiliation[c]{Department of Physics and Astronomy,\\ 
Uppsala University,
Box 516, 75120 Uppsala, Sweden}
\affiliation[d]{Nordita, Stockholm University and KTH Royal Institute of Technology,\\
Hannes Alfv\'ens v\"ag 12,
106 91 Stockholm, Sweden}

\emailAdd{andrea.campoleoni@umons.ac.be}
\emailAdd{arnaud.delfante@umons.ac.be}
\emailAdd{dario.francia@uniroma3.it}
\emailAdd{carlo.heissenberg@physics.uu.se}

\abstract{We study the renormalized action and the renormalized presymplectic potential for Maxwell fields on Anti de Sitter backgrounds of any dimensions. We then use these results to explicitly derive finite boundary charges for angle-dependent asymptotic symmetries. We consider both Poincar\'e and Bondi coordinates, the former allowing us to control the systematics for arbitrary $D$, the latter being better suited for a smooth flat limit.}

\begin{document} 
 
\hfill UUITP-20/23

\maketitle
\flushbottom

\section{Introduction}\label{sec:intro}

The asymptotic symmetries of four-dimensional asymptotically flat gravity and of Yang-Mills theories on Minkowski backgrounds have been related to soft theorems and memory effects, see e.g.~\cite{Strominger:2017zoo} for a review. 
The relation with soft theorems was also extended to gauge theories describing free particles of arbitrary spin and $p$-forms \cite{Campoleoni:2017mbt, Afshar:2018apx, Campiglia:2018see, Francia:2018jtb, Henneaux:2018mgn}, thus presenting itself as a general feature of any gauge theory. 
This connection motivated a reassessment of the asymptotic symmetries of both gravity and gauge theories in spacetimes of dimension greater than four, since proper generalizations of the Bondi-Metzner-Sachs (BMS) group  \cite{Bondi:1962px, Sachs:1962wk, Sachs:1962zza} and of its counterparts in gauge theories \cite{Strominger:2013lka, Barnich:2013sxa, He:2014cra, Campiglia:2015qka,Kapec:2015ena} seem instrumental in establishing the link with soft theorems, that in their turn take the same form in any spacetime dimension.\footnote{See however \cite{Pano:2023slc} for an alternative perspective in $D>4$.}
This observation appears however in tension with the option of defining boundary conditions in higher-dimensional spacetimes 
which, on the one hand, are compatible with radiative solutions while, on the other hand, 
lead to finite-dimensional asymptotic symmetry groups  \cite{Hollands:2003ie, Tanabe:2011es, Hollands:2016oma, Campoleoni:2017qot}.

This led to introduce less restrictive boundary conditions allowing to recover, e.g., the BMS asymptotic group also for spacetime dimensions $D > 4$ \cite{Kapec:2015vwa}, see also \cite{Pate:2017fgt, Aggarwal:2018ilg, Capone:2019aiy, Colferai:2020rte, Capone:2021ouo, Kapec:2021eug, Chowdhury:2022nus, Chowdhury:2022gib, Capone:2023roc}.
Similar infinite-dimensional enhancements of the asymptotic symmetries at null infinity have been identified also for Yang-Mills \cite{Kapec:2014zla, He:2019pll, He:2019ywq, Campoleoni:2019ptc, He:2023bvv} and higher-spin gauge theories \cite{Campoleoni:2020ejn}. 
Most of these works also confirmed the expected link between infinite-dimensional symmetries and soft theorems.
Boundary conditions allowing for infinite-dimensional asymptotic symmetry groups in higher dimensions were discussed at spatial infinity \cite{Esmaeili:2019hom, Henneaux:2019yqq, Fuentealba:2021yvo, Fuentealba:2022yqt, Fuentealba:2023huv} and on (Anti) de Sitter backgrounds \cite{Esmaeili:2019mbw,Fiorucci:2020xto,  Esmaeili:2021szb} too.

While relaxing the boundary conditions so as to allow for a larger class of residual gauge transformations is always possible, selecting bona fide asymptotic symmetries requires a careful analysis of the associated boundary charges, which are the tool allowing one to distinguish between proper and large gauge transformations.  
 Indeed, the boundary charges associated to infinite-dimensional residual symmetries at face value are found to diverge when $D > 4$. 
These divergences can however be renormalized with various techniques, see e.g.~\cite{Henningson:1998gx, Taylor:2000xw,deHaro:2000vlm, Bianchi:2001kw, Skenderis:2002wp,  Papadimitriou:2004ap, Papadimitriou:2004rz, Hollands:2005ya,Papadimitriou:2005ii, Mann:2005yr, Compere:2008us, Papadimitriou:2010as, Compere:2018ylh, Freidel:2019ohg, Compere:2019bua, Compere:2020lrt, Chandrasekaran:2021vyu, Geiller:2022vto, McNees:2023tus}, and in this paper we  systematically explore two of them in the instructive and yet relatively handy example of Maxwell fields in Anti de Sitter (AdS) backgrounds of any dimensions. These investigations can be relevant for both AdS/CFT
and flat holography, as well as to prepare for a corresponding study of de Sitter spacetime, although in the latter case the boundary analysis would require a different interpretation.

In particular, we derive finite boundary charges parameterizing infinite-dimensional asymptotic symmetries in two ways. We first derive them starting from the renormalized bulk action computed via holographic renormalization \cite{deHaro:2000vlm, Bianchi:2001kw, Skenderis:2002wp, Papadimitriou:2005ii}. We then compare the results with those obtained by directly renormalizing the presymplectic potential derived from the naive bulk action \cite{Papadimitriou:2005ii, Freidel:2019ohg}. 
The advantage of the first procedure is that it can be performed in a gauge invariant way at each step and that it provides a complete action principle, including the appropriate boundary contributions. The second approach requires instead a gauge fixing, but it is somewhat simpler to implement. 

We explore and contrast these two complementary renormalization techniques in two different sets of coordinates. We first deal with Poincar\'e coordinates, which provide an established arena for holography and allow one to get results applying to arbitrary $D$, but which do not cover the full AdS space in Lorentzian signature and do not seem appropriate for taking a limit in which the AdS radius $\ell$ tends to infinity. We then move to Bondi coordinates, which are global coordinates and allow for a straightforward flat, $\ell \to \infty$ limit. We compute renormalized actions and presymplectic potentials in both coordinate systems, and study the diffeomorphism connecting the two.
We stress that holographic renormalization is diffeormorphism invariant, which guarantees the agreement between the charges evaluated in Poincar\'e and in Bondi coordinates.

However, since both Bondi and Poincar\'e coordinates represent standard choices in the literature, we opt to present them in parallel, in order to stress their similarities and differences for practical computations.
Our renormalized charges are not conserved, and  we explicitly discuss the associated flux across the spacetime boundary \cite{Compere:2019bua, Fiorucci:2020xto}. Compared to previous analyses of holographic renormalization and boundary charges of Maxwell fields in AdS$_D$ \cite{Taylor:2000xw, Esmaeili:2019mbw, Esmaeili:2021szb}, we provide a general expression for the renormalized action in any $D$ in Poincar\'e cordinates and, thanks to our analysis in Bondi coordinates, we set up a framework that allows one to recover the surface charges at null infinity \cite{Freidel:2019ohg}, rather than spatial infinity, in a smooth flat limit. 
Let us also mention that our solution spaces include logarithms of the radial coordinate in odd spacetime dimensions.

A natural extension of the current work will be to apply the same strategy and the tools here developed to any linear gauge theory. On the one hand, this is expected to allow one to generalize the boundary conditions employed in \cite{Campoleoni:2016uwr} to derive conserved asymptotic charges for higher-spin fields in AdS. In addition, the option to get charges at null infinity via the $\ell \to \infty$ limit computed in Bondi coordinates  should provide an alternative derivation of the boundary charges associated to the infinite-dimensional higher-spin asymptotic symmetries of \cite{Campoleoni:2020ejn}, that played a crucial role in establishing the link with soft theorems.

In order to display our strategy in a simpler setting, we begin in section~\ref{sec: Spin-0} by evaluating the renormalized action and presymplectic potential for a massless scalar field. In section~\ref{sec:Spin1Poincare} we then  move to Maxwell fields on AdS backgrounds, in Poincar\'e coordinates. This allows us to obtain results applying to generic spacetime dimensions, while in section~\ref{sec:spin1_bondi} we recompute from scratch the renormalized action and presymplectic potential in Bondi coordinates in a number of examples. In section~\ref{sec:further} we eventually discuss how one can derive the charges in Bondi coordinates by starting from the more general results obtained in Poincar\'e coordinates and we also discuss how to take the flat limit. A couple of appendices recall some useful facts about the geometry of AdS and the key features of the covariant phase space formalism that we use to renormalize the presymplectic potential.

\section{Scalar fields} \label{sec: Spin-0}

\subsection{Poincar\'e coordinates}  \label{sec: Spin-0 FG}
As a warm up to illustrate the renormalization techniques, let us begin by discussing the case of a free massless scalar field. In this section, we employ Poincar\'e coordinates:
\begin{equation} \label{AdS FG Bulk Metric}
\text{d}s^2 = \frac{1}{z^2} \left( \ell^2 \text{d}z^2 + \eta_{ab}\,\text{d}x^a \text{d}x^b \right) ,
\end{equation}
where $\eta_{ab}$ is the Minkowski metric and $a,b \in \{0,\ldots,D-2\}$ (see also appendix~\ref{app:Poincare}).\footnote{Although we restrict for simplicity to the Minkowski metric $\eta_{ab}$ on the boundary, we emphasize that our results also hold for a generic flat boundary metric.}

\paragraph{Solution space}

We consider the action of a free massless scalar field,
\begin{equation} \label{Action of spin-0 massless field}
S = - \frac{1}{2} \, \int \text{d}^{D}x \, \sqrt{-g} \, \nabla^\mu \Phi \, \nabla_{\!\mu} \Phi \, ,
\end{equation}
whose equation of motion
\begin{equation} \label{eom of spin-0 massless field}
    \partial_\mu \left( \sqrt{-g} \, \partial^\mu \Phi \right) = 0 
\end{equation}
in Poincar\'e coordinates looks
\begin{equation}
    z \, \ell^2 \, \Box \Phi + \left( z \, \partial_z - (D-2) \right) \partial_z \Phi = 0 \, ,
\end{equation}
where $\Box := \eta^{ab} \partial_a \partial_b $. Considering an asymptotic expansion of the field in the radial $z$-coordinate of the form
\begin{equation} \label{generic radial expansion of the scalar}
    \Phi(z,x^a) = \sum_{n \geq 0} z^n \, \phi^{(n)}(x^a) + \sum_{n \geq 0} z^n \, \log z \, \widetilde{\phi}^{(n)}(x^a) \, ,
\end{equation}
where we are excluding solutions that explode at the boundary $z \to 0$, 
one finds the recursion relations 
\begin{subequations}
\begin{align}
    &\Box \phi^{(n-2)} + \frac{1}{\ell^2} \, n \, (n - D + 1) \, \phi^{(n)} + \frac{1}{\ell^2} \, (2 n - D + 1) \, \widetilde{\phi}^{(n)} = 0 \, ,\\[5pt]
    &\Box \widetilde{\phi}^{(n-2)} + \frac{1}{\ell^2} \, n \, (n - D + 1) \, \widetilde{\phi}^{(n)} = 0 \, .
\end{align}
\end{subequations}
These, in their turn, determine the asymptotic solution space, where $\phi^{(0)}$ and $\phi^{(D-1)}$ are arbitrary functions of the boundary coordinates $x^a$, while the coefficients that will be relevant in the following read
\begin{subequations} \label{solution of spin-0 massless field}
\begin{align}
    &\phi^{(2n)} = \frac{\ell^{2 n} \, (-4)^{-n} \, \Gamma \left(\frac{3-D}{2}\right) \Box^n \phi^{(0)}}{\Gamma (n+1) \, \Gamma
   \left(n-\frac{D-3}{2}\right)} \, , \qquad \phi^{(2n+1)} = 0 \, , \qquad \left(0 < n < \frac{D-1}{2}\right),\\
    &\widetilde{\phi}^{(D-1)} = - \frac{\ell^2}{D-1} \, \Box \phi^{(D-3)} \,,
\end{align}
\end{subequations}
where the last equation is trivial for even $D$.
Therefore, the asymptotic radial expansion is, in even $D$ 
\begin{equation} \label{even expansion of spin-0 massless field}
    \Phi(z,x^a) = \sum_{n \geq 0} z^{2n} \, \phi^{(2n)}(x^a) + \sum_{n \geq 0} z^{2n+D-1} \, \phi^{(2n+D-1)}(x^a) \, ,
\end{equation}
and in odd $D$ 
\begin{equation} \label{odd expansion of spin-0 massless field}
    \Phi(z,x^a) = \sum_{n \geq 0} z^{2n} \, \phi^{(2n)}(x^a) + \log z \, \sum_{n \geq 0} z^{2n+D-1} \, \widetilde{\phi}^{(2n+D-1)}(x^a) \, .
\end{equation}

As noticed in \cite{deHaro:2000vlm,Skenderis:2000in,Bianchi:2001kw,Skenderis:2002wp} for (massive) scalars in any $D$, the on-shell action typically diverges due to contributions localized at the boundary ($z \to 0$), and one has to implement holographic renormalization to set up the variational principle.  

\paragraph{Holographic renormalization}

We start by computing the regularized on-shell action, 
\begin{equation}
S_{\text{reg}}^{(\epsilon)} = - \frac{1}{2} \int_{z \geq \epsilon} \text{d}^{D}x \, \sqrt{-g} \, \mathcal F_\mu \, g^{\mu\nu}\,\mathcal F_\nu \, ,
\end{equation}
where we defined 
\begin{equation}
    \mathcal{F}_\mu := \nabla_{\!\mu} \Phi \, ,
\end{equation}
in order to mimic the role played by the field strength in the spin-one case. Since the boundary is now located at $z-\epsilon=0$, its normal is given by $n_\mu = \partial_\mu(z-\epsilon)= \delta_\mu^z$, so that  
\begin{equation}
    \sqrt{-g}\,n_\mu  g^{\mu\nu}\,\mathcal F_\nu = \frac{1}{\ell z^{D-2}}\,\mathcal F_z\, .
\end{equation}
By Stokes' theorem we can then write the regulated on-shell action as ($\approx$ stands for on-shell equality)
\begin{equation} \label{regulated on-shell action of spin-0 massless field}
    S_{\text{reg}}^{(\epsilon)} \approx \frac{1}{2 \ell \epsilon^{D-2}} \int_{z=\epsilon} \text{d}^{D-1}x \, \Phi \, \mathcal{F}_z \: ,
\end{equation}
where the global sign reflects the fact that $z=\epsilon$ is the lower limit of the integration. To leading order for small $z$, the field $\Phi$ is $z$-independent, while the ``field-strength'' component  $\mathcal F_z$ scales like $z$. Therefore, the regulated action scales like $\epsilon^{3-D}$
and, whenever $D\ge4$, it has divergent contributions. We will now see in a few examples how can one subtract those divergences.

Let us consider $D=6$. In this case the free functions in the asymptotic expansion \eqref{generic radial expansion of the scalar} are $\phi^{(0)}$ and $\phi^{(5)}$.  The regulated on-shell action \eqref{regulated on-shell action of spin-0 massless field} reads
\begin{equation}\label{SregulatedScalar214Bas}
S_{\text{reg}}^{(\epsilon)} = \int_{z=\epsilon} \text{d}^5x \, \Phi \,\Psi \,,\qquad
\Psi =\frac{1}{2\ell\,\epsilon^{4}}\,\mathcal{F}_z\,,
\end{equation}
where $\Psi$ diverges like $1/\epsilon^3$ to leading order as $\epsilon\to0$.
Our strategy will be to find a counterterm action $S_{\text{ct}}^{(\epsilon)}$ such that
\begin{equation}
S_{\text{sub}}^{(\epsilon)}
=
S_{\text{reg}}^{(\epsilon)} + S_{\text{ct}}^{(\epsilon)}  =  \int_{z=\epsilon} \text{d}^5x \, \Phi \,\tilde\Psi
\end{equation}
and $\tilde\Psi$ is finite as $\epsilon\to0$. Since the additional $\Phi$ appearing in the action is finite at the boundary, achieving this will be enough to produce a finite subtracted action.
With this objective in mind, we thus proceed by writing
\begin{equation}\label{SregulatedScalar214}
S_{\text{reg}}^{(\epsilon)} = \int_{z=\epsilon} \frac{\text{d}^5x}{2 \ell} \, \Phi \left( \epsilon^{-3} F_z^{(1)} + \epsilon^{-1} F_z^{(3)}\right) + \mathcal{O}(1) \,,
\end{equation}
where $F_z^{(k)}$ denotes the coefficient of $z^k$ in the expansion of $\mathcal{F}_z (z, x^a)$.
Following the above strategy, since $\Phi$ remains finite as $\epsilon\to0$ and thus cannot introduce additional divergences,we will eventually choose to not expand it. This has the advantage of simplifying the intermediate calculations, and will make it possible to obtain results holding for generic $D$ below. A similar choice in the spin-one case will allow us to express all counterterms directly in terms of the field strength. 

More explicitly, expressing the $F_z^{(k)}$ in \eqref{SregulatedScalar214} in terms of $\phi^{(0)}$ by using the equations of motion, expanding also the $\Phi$ appearing in \eqref{SregulatedScalar214} and multiplying its expansion with the one in the round parenthesis, we obtain the following expression for the divergent terms expressed in terms of $\phi^{(k)}$, 
\begin{equation}\label{explicit-check}
S_{\text{reg}}^{(\epsilon)} 
=
\frac{1}{\ell} \int d^5x \left[
\frac{1}{\epsilon^3}  \phi^{(0)} \phi^{(2)} +\frac{1}{\epsilon} \left(
2\phi^{(0)} \phi^{(4)}+(\phi^{(2)})^2
\right)
\right]
+\mathcal O(1)\,,
\end{equation}
which shows the dependence on the counterterm from the boundary data. However,
inverting the expansion \eqref{even expansion of spin-0 massless field}, 
we can express back the regulated action in terms of the bulk field, finding
\begin{equation}\label{SregExample}
S_{\text{reg}}^{(\epsilon)} = \int_{z=\epsilon} \text{d}^5x \, \frac{\ell}{6 \epsilon^3} \, \Phi \left( 1 + \frac{\ell^2 \epsilon^2}{3} \Box \right) \partial \cdot \mathcal{F} + \mathcal{O}(1)\,.
\end{equation}
Let us now note that the operation of expanding $\Phi$ in terms of its asymptotic expansion coefficients $\phi^{(k)}$ and then inverting this expansion to re-express the resulting sum in terms of $\Phi$ acts as the identity on $\Phi$, i.e.~gives back $\Phi$ itself (regardless whether or not one first multiplies this expansion with that of $\mathcal F_z$ in \eqref{SregulatedScalar214}).
This fact can be checked explicitly using \eqref{explicit-check} for the present example, but holds in any dimension.
Therefore, while needed in order to obtain the explicit expression \eqref{explicit-check} of $S_{\text{reg}}^{(\epsilon)}$ in terms of $\phi^{(k)}$, explicitly expanding $\Phi$ is not necessary in order to arrive at the form \eqref{SregExample} for  $S_{\text{reg}}^{(\epsilon)}$ in terms of the bulk field, which is what we need to the goal of computing the presymplectic potential.
In the following, we will restrict our attention to expressions in terms of the bulk fields, so that we will not need to expand the overall $\Phi$ appearing in the regulated action, which considerably simplifies the intermediate steps especially as $D$ increases. 
We can thus define the counterterm action as follows in the present case,
\begin{equation}
    S_{\text{ct}}^{(\epsilon)} = - \int_{z=\epsilon} \text{d}^5x \, \frac{\ell}{6 \epsilon^3} \, \Phi \left( 1 + \frac{\ell^2 \epsilon^2}{3} \Box \right) \partial \cdot \mathcal{F}
\end{equation} 
and
write the subtracted action as
\begin{equation} \label{subtracted0}
\begin{split}
S_{\text{sub}}^{(\epsilon)} &= S_{\text{reg}}^{(\epsilon)} + S_{\text{ct}}^{(\epsilon)} \\
&= - \int_{z > \epsilon} \text{d}^6x \, \frac{\sqrt{-g}}{2} \, \mathcal{F}^\mu \, \mathcal{F}_\mu + \int_{z=\epsilon} \text{d}^5x \, \frac{\ell}{6 \epsilon^3} \, \mathcal{F}^a \left( 1 + \frac{\ell^2 \epsilon^2}{3} \Box \right) \mathcal{F}_a + \mathcal{O}(\epsilon) \, .
\end{split}
\end{equation}
Here and in the following, the indices $a,b,\dots$ are lowered and raised respectively via the metric $\eta_{ab}$ and its inverse $\eta^{ab}$.
In \eqref{subtracted0}, we have assumed $\mathcal F_a$ vanishes on the boundary of the $z=\epsilon$ surface (which allows one to freely integrate by parts w.r.t. $x^a$); alternatively we could add the extra ``corner term''
$    S_{\text{corner}}^{(\epsilon)} = \int_{z=\epsilon} \text{d}^5x \, \partial_a \left[ \frac{\ell}{6 \epsilon^3} \, \Phi \left( 1 + \frac{\ell^2 \epsilon^2}{3} \Box \right) \mathcal{F}^a \right] ,
$
which has the effect of canceling such integrations by parts, and clearly does not spoil the variational principle.
By varying it on-shell and by injecting the asymptotic solution, we obtain
\begin{equation}
\begin{split}
\delta S_{\text{sub}}^{(\epsilon)} &\approx \int_{z=\epsilon} \text{d}^5 x \left[ \frac{1}{\ell \epsilon^4} \, \delta \Phi \, \mathcal{F}_z - \frac{\ell}{3 \epsilon^3} \, \delta \Phi \left( 1 + \frac{\ell^2 \epsilon^2}{3} \Box \right) \partial \cdot \mathcal{F} \right] + \mathcal{O}(\epsilon)\\
&= \int_{z=\epsilon} \text{d}^5x \, \frac{1}{\ell} \, \delta \Phi \, F_z^{(4)} + \mathcal{O}(\epsilon) \, ,
\end{split}
\end{equation}
thus leading to 
\begin{align}
\delta S_{\text{ren}} := \lim_{\epsilon \to 0} \delta S_{\text{sub}}^{(\epsilon)} \approx \int  \frac{\text{d}^5x}{\ell} \, \delta \Phi \, F_z^{(4)} \, .
\end{align}

In arbitrary even $D\geq4$, by expressing the boundary contributions encoded into the $F^{(n)}_z$ in terms of bulk field $\Phi$, gathering the divergent pieces and assuming the same falloff for $\mathcal F_a$ (equivalently, adding a suitable corner term), one obtains the counterterm action
\begin{equation}\label{counterterms1evenDspin0}
    S_{\text{ct}}^{(\epsilon)} = \frac{{\ell}}{2 (D-3) \epsilon^{D-3}} \int_{z=\epsilon} \text{d}^{D-1} x \, \mathcal{F}^{a} \left( 1 + \sum_{k=1}^{\frac{D-4}{2}} {\ell^{2k}} \epsilon^{2k} c_N^{(k+1)} \Box^k \right) \mathcal{F}_{a} \, ,
\end{equation}
where the coefficients are fixed recursively by
\begin{equation} \label{spin1 FG variational generic recursive coefficient}
    c_N^{(k)} = - \frac{ (-4)^{-k+1} (k-1) \Gamma\left(\frac{-D+3}{2}\right) }{ \Gamma\left(k\right) \Gamma\left(\frac{2k-D+3}{2}\right) } - \sum_{q=0}^{k-3} \left( c_N^{(q+2)} \frac{ (-4)^{-k+2+q} \Gamma\left(\frac{-D+3}{2}\right) }{ \Gamma\left({k-1-q}\right) \Gamma\left(\frac{2k-D+1-2q}{2}\right) } \right) .
\end{equation}
Let us remark that the expression \eqref{counterterms1evenDspin0} for the counterterm Lagrangian can be rewritten in the following way,
\begin{equation}\label{diffinvEvenDs0}
S_{\text{ct}}^{(\epsilon)} = \frac{{\ell}}{2 (D-3)} \int_{z=\epsilon} \text{d}^{D-1} x\, \sqrt{-\gamma} \, \gamma^{ab}\,\mathcal{F}_{a} \left( 1 + \sum_{k=1}^{\frac{D-4}{2}} {\ell^{2k}} c_N^{(k+1)} \Box_\gamma^k \right) \mathcal{F}_{b} \, ,
\end{equation}
where $\gamma_{ab} = \eta_{ab}/\epsilon^2$ is the induced metric on the surface $z=\epsilon$ and $\Box_\gamma = \gamma^{ab}\partial_a\partial_b$. This makes the diffeomorphism invariance of the procedure manifest, since no explicit (coordinate-dependent) dependence on $\epsilon$ appears in \eqref{diffinvEvenDs0}.
The on-shell variation of the subtracted action takes the form
\begin{equation}
\delta S_{\text{sub}}^{(\epsilon)} \approx \int_{z=\epsilon} \text{d}^{D-1} x \, \delta \Phi \left[ \frac{\epsilon^{2-D}\mathcal{F}_{z}}{2 \ell} - \frac{\ell\,\epsilon^{3-D}}{D-3} \left( 1 + \sum_{k=1}^{\frac{D-4}{2}} {\ell^{2k}} \epsilon^{2k} c_N^{(k+1)} \Box^k \right) \partial \cdot \mathcal{F} \right],
\end{equation}
and yields the following variation of the renormalized on-shell action 
\begin{equation} \label{spin1 FG variational generic renormalized on-shell action}
\delta S_{\text{ren}} = \lim_{\epsilon \to 0} \delta S_{\text{sub}}^{(\epsilon)} \approx \frac{1}{{\ell}} \int \text{d}^{D-1} x \,  \delta \Phi \, F_{z}^{(D-2)} \, ,
\end{equation}
depending only on the subleading free function since $F_z^{(D-2)} = (D-1) \phi^{(D-1)}$.

In the odd-dimensional case there appear new finite terms in the counterterm action when expressed in terms of bulk fields, although, in the present renormalization scheme, we systematically drop them. For concreteness, let us first illustrate this phenomenon for the case $D=7$. The asymptotic expansion of the scalar field is given by \eqref{odd expansion of spin-0 massless field} where the arbitrary orders are $\phi^{(0)}$ and $\phi^{(6)}$. As explained below \eqref{SregulatedScalar214Bas}, if we don't radially expand 
$\Phi$, the regulated on-shell action is 
\begin{equation}\label{explicitD7}
S_{\text{reg}}^{(\epsilon)} = \int_{z=\epsilon} \frac{\text{d}^6x}{2 \ell} \, \Phi \left( \epsilon^{-4} F_z^{(1)} + \epsilon^{-2} F_z^{(3)} + \log \epsilon \widetilde{F}_z^{(5)} 
\right) + 
\mathcal O(1)
\end{equation}
and re-expressing it in terms of the bulk fields leads to
\begin{equation}
    S_{\text{reg}}^{(\epsilon)} = \int_{z=\epsilon} \text{d}^6x \frac{\ell}{8 \epsilon^4} \Phi \left( 1 + \frac{\ell^2 \epsilon^2}{8} \Box - \frac{\ell^4 \epsilon^4}{16} \log \epsilon \, \Box^2 \right) \partial^a\mathcal{F}_a  
    + \mathcal{O}(1) 
    \, .
\end{equation}
As anticipated, in this last step additional contributions of the type $\epsilon^4 \ell^4 \Box^2$ would appear inside the round parenthesis, but we systematically drop them as they correspond to finite local functionals of $\phi^{(0)}$, and, following \cite{deHaro:2000vlm,Skenderis:2000in,Bianchi:2001kw,Skenderis:2002wp}, we recall that the result after the subtraction of the divergent parts is only defined up to the addition of such scheme-dependent finite terms. An additional source of (harmless) ambiguities of this type would come from first expanding the overall $\Phi$ in \eqref{explicitD7}, dropping the resulting finite pieces depending on $\phi^{(k)}$ and then re-expressing the divergent contributions in terms of the bulk field $\Phi$.
The counterterm action, with the inclusion of the corner term, thus takes the following form:
\begin{equation}
    S_{\text{ct}}^{(\epsilon)} = \int_{z=\epsilon} \text{d}^6x \frac{\ell}{8 \epsilon^4} \mathcal{F}^a \left( 1 + \frac{\ell^2 \epsilon^2}{8} \Box - \frac{\ell^4 \epsilon^4}{16} \log \epsilon \, \Box^2 \right) \mathcal{F}_a  
\end{equation}
and, in the limit $\epsilon \to 0$, the on-shell variation of the subtracted action $S_{\text{sub}}^{(\epsilon)} = S_{\text{reg}}^{(\epsilon)} + S_{\text{ct}}^{(\epsilon)}$ yields
\begin{equation}
    \delta S_{\text{ren}} \approx \frac{1}{\ell} \int \text{d}^6x \, \delta \Phi \, F_z^{(5)} \, .
\end{equation}

Similarly to the even-dimensional case, we can generalize the procedure to any odd dimension, for $D > 4$, by means of  the following counterterm 
\begin{equation}\label{countertermss0oddD}
    S_{\text{ct}}^{(\epsilon)} = \int_{z=\epsilon} \frac{\ell \,\text{d}^{D-1} x}{2 (D-3) \epsilon^{D-3}} \mathcal{F}^{a} \left( 1 + \sum_{k=1}^{\frac{D-5}{2}} {(\ell \epsilon)^{2k}} c_N^{(k+1)} \Box^k - {(\ell \epsilon)^{D-3}} \log \epsilon \, c_L \Box^{\frac{D-3}{2}} \right) \mathcal{F}_{a} \, ,
\end{equation}
which, as usual, includes a corner-term contribution.
The last two equations involve the recursive coefficients $c_N^{(k)}$ \eqref{spin1 FG variational generic recursive coefficient} as well as a logarithmic coefficient
\begin{equation}
    c_L = 
    {\frac{2^{4-D}}{\Gamma\left(\frac{D-1}{2}\right) \Gamma\left(\frac{D-3}{2}\right)}}
    .
\end{equation}
Let us rewrite \eqref{countertermss0oddD} also in a manner analogous to \eqref{diffinvEvenDs0}, 
\begin{equation}\label{countertermss0oddDalmostdiff}
    S_{\text{ct}}^{(\epsilon)} = \int_{z=\epsilon} \frac{\ell \,\text{d}^{D-1} x}{2 (D-3)}\,\sqrt{-\gamma}\,\gamma^{ab} \mathcal{F}_{a} \left( 1 + \sum_{k=1}^{\frac{D-5}{2}} {\ell^{2k}} c_N^{(k+1)} \Box_\gamma^k - {\ell^{D-3}} \log \epsilon \, c_L \Box_\gamma^{\frac{D-3}{2}} \right) \mathcal{F}_{b} \, ,
\end{equation}
where all the dependence on $\epsilon$ enters via the cutoff surface and the induced metric $\gamma_{ab}$ thereon, except for the $\log\epsilon$ term.
When one approaches the boundary, it leads to the same expression for the variation of the renormalized on-shell action as in \eqref{spin1 FG variational generic renormalized on-shell action}.

\paragraph{Symplectic renormalization}

Using the covariant phase space formalism (see appendix \ref{app:CPS}) one can renormalize the scalar theory directly at the level of the symplectic structure \cite{Papadimitriou:2005ii, Freidel:2019ohg}. As we shall see, an advantage of this set-up is that we will obtain a radial renormalization equation providing the systematics for the introduction of the additional terms.

In order to see this explicitly, following \cite{Freidel:2019ohg}, let us write the variation of the action \eqref{Action of spin-0 massless field} in the form
\begin{align}
    \delta \mathscr{L} &= (\text{eom}) \, \delta \Phi + \partial_\mu \Theta^\mu \, , \label{Spin-0 FG Presymp Radial Eq To take first}
\end{align}
where the presymplectic potential is
\begin{equation} \label{presymplectic potential massless scalar field strength}
    \Theta^\mu = - \sqrt{-g} \, \mathcal{F}^\mu \, \delta \Phi \: .
\end{equation}
Splitting $\Theta^\mu$ into radial and boundary components and factoring out an overall $z^{-(D-2)}$ for convenience, we define
\begin{align}
    \Theta^z & = z^{-(D-2)} \, \widetilde{\Theta}^z \, , \qquad \widetilde{\Theta}^z = - \frac{1}{\ell} \, \mathcal{F}_z \, \delta \Phi \, , \label{expression of Theta z in FG spin-0 massless field}\\[1ex]
    \Theta^a & = z^{-(D-2)} \, \widetilde{\Theta}^a \, , \qquad \widetilde{\Theta}^a = - \ell \, \mathcal{F}^a \, \delta \Phi \, ,\label{expression of Theta a in FG spin-0 massless field}
\end{align}
and
\begin{equation} \label{expression of mathscr L in FG spin-0 massless field}
    \mathscr{L}  = z^{-(D-2)} \, \widetilde{\mathscr{L}} \, , \qquad \widetilde{\mathscr{L}} = -  \frac{1}{2\ell} \Big( ( \mathcal{F}_z )^2 + \ell^2 \, \mathcal{F}^a \, \mathcal{F}_a \Big) 
\end{equation}
for the off-shell Lagrangian.
Thus, we can split the total divergence in \eqref{Spin-0 FG Presymp Radial Eq To take first} into radial and boundary divergences and obtain the asymptotic renormalization equation: 
\begin{equation}
    \frac{1}{z} \Big(z \, \partial_z - (D-2) \Big) \widetilde{\Theta}^z \approx  \delta \widetilde{\mathscr{L}} - \partial_a \widetilde{\Theta}^a \, .\label{Spin-0 FG Presymp Radial Eq}
\end{equation}
The radial equation \eqref{Spin-0 FG Presymp Radial Eq} implies that the symplectic potential can be made finite on-shell by subtracting counterterms which are either total variations or total boundary derivatives, as allowed by the ambiguity equation \eqref{ambiguities def}. Indeed, assuming an asymptotic radial expansion of the form
\begin{equation} \label{generic radial expansion of the scalar Theta and L}
    \widetilde{\Theta}^\mu = \sum_n z^n \left( \widetilde{\Theta}^{\mu}_{(n)} + \log z \, \widetilde{\theta}^{\mu}_{(n)} \right) , \qquad \widetilde{\mathscr{L}} = \sum_n z^n \left( \widetilde{\mathscr{L}}^{(n)} + \log z \, \widetilde{\ell}^{(n)} \right) ,
\end{equation}
and substituting in \eqref{Spin-0 FG Presymp Radial Eq} one finds 
\begin{equation}
    (n-D+2) \, \widetilde{\Theta}^{z}_{(n)} + \widetilde{\theta}^{z}_{(n)} \approx \delta \widetilde{\mathscr{L}}^{(n-1)} - \partial_a \widetilde{\Theta}^{a}_{(n-1)} \, , \label{Spin-0 FG Presymp Radial Eq order}
\end{equation}
and similar equations involving the coefficients of the log terms. The above equation shows in particular that the orders $n < D-2$ of $\widetilde{\Theta}^{z}_{(n)}$, i.e.\ the ones that come with divergent prefactors in $\Theta^z$, are fixed on-shell to be total derivatives plus total variations, while $\widetilde{\Theta}^{z}_{(D-2)}$, which gives the finite order of $\Theta^z$, is left undetermined. 
The remaining terms do not contribute in the asymptotic limit $z \to 0$.  Let us illustrate the procedure for the two cases of even and odd dimensions.

For even $D \ge 4$, we can write
\begin{equation}
    \Theta^z = \sum_{n=0}^{\frac{D-4}{2}} z^{2n-D+3} \, \widetilde{\Theta}^{z}_{(2n+1)} + \widetilde{\Theta}^{z}_{(D-2)} + \mathcal{O}(z) \, ,
\end{equation}
where the summation includes all diverging terms, given on-shell by (see \eqref{solution of spin-0 massless field})
\begin{equation} \label{order 2pp1 of the expression of Theta z in FG spin-0 massless field}
    \widetilde{\Theta}^{z}_{(2n+1)} = - \frac{1}{\ell} \, \sum_{q=0}^{p} F_{z}^{(2(n-q)+1)} \, \delta \phi^{(2q)} \, ,
\end{equation}
while the finite order is
\begin{equation}
    \widetilde{\Theta}^{z}_{(D-2)} = - \frac{1}{\ell} \, F_{z}^{(D-2)} \, \delta \phi^{(0)} \, .
\end{equation}
According to the radial equation \eqref{Spin-0 FG Presymp Radial Eq order}, we can cancel the divergent orders adding terms
of the form $\partial_a \widetilde{\Theta}^{a}_{(2n)}$ and $\delta \widetilde{\mathscr{L}}^{(2n)}$, with
\begin{equation} \label{order 2p of the expression of Theta a in FG spin-0 massless field}
    \widetilde{\Theta}^{a}_{(2n)} = - \ell \sum_{q=0}^{2n-1} F^a_{(2n-2q)} \, \delta \phi^{(2q)} \, ,
\end{equation}
and
\begin{subequations} \label{order 2p of the expression of L in FG spin-0 massless field}
    \begin{align}
        \widetilde{\mathscr{L}}^{(4n)} &= - \frac{1}{2\ell} \left[ \left( F_z^{(2n)} \right)^2  + \ell^2 \, \sum_{q=0}^{4n-1} F^a_{(4n-2q)} \, F_a^{(2q)} \right] ,\\
        \widetilde{\mathscr{L}}^{(4n+2)} &= - \frac{1}{2\ell} \left[ \sum_{q=0}^{2n} F_z^{(2(2n-q)+1)} \, F_z^{(2q+1)} + \ell^2 \, \sum_{q=0}^{4n+1} F^a_{(2(2n-q+1))} \, F_a^{(2q)} \right] .
    \end{align}
\end{subequations}
The counterterm is therefore
\begin{subequations}
    \begin{align}
        \Theta^z_{\text{ct}} &= \delta B_{\text{ct}} - \partial_a C^a_{\text{ct}} \, ,\\
        B_{\text{ct}} &= \sum_{p=0}^{\frac{D-4}{2}} \frac{z^{-(D-2n-3)}}{D-2n-3} \widetilde{\mathscr{L}}^{(2n)} \, , \qquad C^a_{\text{ct}} = - \sum_{p=0}^{\frac{D-4}{2}} \frac{z^{-(D-2n-3)}}{D-2n-3} \widetilde{\Theta}^a_{(2n)} \, ,
    \end{align}
\end{subequations}
while the renormalized presymplectic potential is
\begin{equation} \label{renorm presymp scalar FG generic}
    \Theta^z_{\text{ren}} = \lim_{z\to 0} \left( \Theta^z + \Theta^z_{\text{ct}} \right) = \widetilde{\Theta}^{z}_{(D-2)} = - \frac{1}{\ell} \, F_{z}^{(D-2)} \, \delta \phi^{(0)} \, .
\end{equation}

By following similar steps, we can deal with the case of odd $D > 4$. The asymptotic expansion of the radial component of the presymplectic potential is
\begin{equation}
    \Theta^z = \sum_{p=0}^{\frac{D-5}{2}} z^{2n-D+3} \, \widetilde{\Theta}^{z}_{(2n+1)} + \log z \, \widetilde{\theta}_z^{(D-2)} + \widetilde{\Theta}^{z}_{(D-2)} + \mathcal{O}(z^2) \, ,
\end{equation}
where the diverging orders are the terms in the sum and the log term. They are given on-shell respectively by \eqref{order 2pp1 of the expression of Theta z in FG spin-0 massless field} and
\begin{equation}
    \widetilde{\theta}^{z}_{(D-2)} = - \frac{1}{\ell} \, \widetilde{F}_z^{(D-2)} \, \delta \phi^{(0)} \, ,
\end{equation}
while the finite order is given by
\begin{equation}
    \widetilde{\Theta}^{z}_{(D-2)} = - \frac{1}{\ell} \, \sum_{n=0}^{\frac{D-3}{2}} F_z^{(D-2-2n)} \, \delta \phi^{(2n)} \, .
\end{equation}
Using  \eqref{Spin-0 FG Presymp Radial Eq order} we can build a counterterm $\Theta^z_{\text{ct}} = \delta B_{\text{ct}} - \partial_a C^a_{\text{ct}}$, where
\begin{align}
        B_{\text{ct}} &= \sum_{n=0}^{\frac{D-5}{2}} \frac{z^{-(D-2n-3)}}{D-2n-3} \widetilde{\mathscr{L}}^{(2n)} - \log z \, \widetilde{\mathscr{L}}^{(D-3)} \, ,\\
        C^a_{\text{ct}} &= - \sum_{n=0}^{\frac{D-5}{2}} \frac{z^{-(D-2n-3)}}{D-2n-3} \widetilde{\Theta}^a_{(2n)} + \log z \, \widetilde{\Theta}^a_{(D-3)} \, .
\end{align}
Therefore, the renormalized presymplectic potential is
\begin{equation}
        \Theta^z_{\text{ren}} = \lim_{z\to 0} \left( \Theta^z + \Theta^z_{\text{ct}} \right) = \widetilde{\Theta}^{z}_{(D-2)} \, .
\end{equation}
Upon further adding the following corner term
\begin{equation}
    \Theta^{\text{corner}}_a = \frac{i \, (-1)^{\frac{i}{4}(i^D + i D)} \, \Gamma \left( \frac{1}{4} (D - i^{D+1}) \right) \Gamma \left( 1 - \frac{D}{4} - \frac{i}{4} e^{\frac{i D \pi}{2}} \right) }{\sqrt{\pi} \, 2^{\frac{D-3}{2}} \ell^{D-2} \Gamma \left( \frac{D-3}{2} \right) \Gamma \left( \frac{D-1}{2} \right) \Gamma \left( \frac{3-D}{2} \right)} \partial_a \Box^{\frac{D-5}{2}} \partial \cdot F^{(0)} \delta \phi^{(0)} \,,
\end{equation}
the above renormalized potential can be expressed in terms of the arbitrary coefficients as 
\begin{equation} \label{Thetazren generic odd FG scalar w ambi}
    \Theta^z_{\text{ren}} = - \frac{1}{\ell} F_z^{(D-2)} \delta \phi^{(0)} - \left(\frac{\ell}{2}\right)^{D-2} \frac{H_{\frac{D-3}{2}}}{\Gamma \left( \frac{D-1}{2} \right)^2} \Box^{\frac{D-3}{2}} \partial \cdot F^{(0)} \delta \phi^{(0)}
\end{equation}
where we denoted the $n$-th harmonic number by $H_n$
\begin{equation}
    H_n = \sum_{k=1}^n \frac{1}{k} \, .
\end{equation}
One can see that the last term of \eqref{Thetazren generic odd FG scalar w ambi} is a combination of symplectic ambiguities \eqref{ambiguities def}. Indeed, it can be cancelled through the addition of
\begin{equation}
    \Theta^{\text{bdy}} = \frac{1}{2} \left(\frac{\ell}{2}\right)^{D-2} \frac{H_{\frac{D-3}{2}}}{\Gamma \left( \frac{D-1}{2} \right)^2} \delta \left( \Box^{\frac{D-3}{2}} F^a_{(0)} F_a^{(0)} \right)
\end{equation}
and
\begin{equation}
    \Theta^{\text{corner}}_a = - \left(\frac{\ell}{2}\right)^{D-2} \frac{H_{\frac{D-3}{2}}}{\Gamma \left( \frac{D-1}{2} \right)^2} \, \Box^{\frac{D-3}{2}} F_a^{(0)} \delta \phi^{(0)} \, .
\end{equation}
At the end, we obtain the same result \eqref{renorm presymp scalar FG generic} as for the even-dimensional case. 
Note that the results obtained for the presymplectic potential from the two renormalization procedures lead to the exact same expression in even dimensions, see \eqref{spin1 FG variational generic renormalized on-shell action} and \eqref{renorm presymp scalar FG generic}, while they differ by local terms in $\phi^{(0)}$ in odd dimensions, see \eqref{spin1 FG variational generic renormalized on-shell action} and \eqref{Thetazren generic odd FG scalar w ambi}. We will encounter a similar situation for the case of spin one.

\subsection{Bondi coordinates} \label{sec: Spin-0 Bondi}

We now revisit the previous construction in Bondi coordinates,
\begin{equation}\label{ds2Bondi}
	\text{d}s^2 = -\left(1+\frac{r^2}{\ell^2}\right)\text{d}u^2-2\text{d}u \text{d}r + r^2\gamma_{ij}\, \text{d}x^i\text{d}x^j\,,
\end{equation}
where $\gamma_{ij}$ is the  metric on the unit $(D-2)$-dimensional celestial sphere, parameterised by the angular coordinates $x^i$ (see also appendix~\ref{app:Bondi}). Furthermore, we focus on the renormalization of the presymplectic potential. The resulting techniques  will be instrumental to consider the flat limit of the surface charges in the spin-one case discussed in section~\ref{sec:further}. 

\paragraph{Solution space}

In Bondi coordinates, in which the AdS metric is given by \eqref{ds2Bondi}, the equation of motion \eqref{eom of spin-0 massless field} for a massless scalar reads 
\begin{equation}
    \left [\Big( \Delta + r^2 \, \partial_r^2 + r \, (D-2) \, \partial_r \Big) - r \Big( 2 \, r \, \partial_r + D - 2 \Big) \partial_u + \frac{r^2}{\ell^2} \Big( r^2 \, \partial_r + D \, r \Big) \partial_r \right] \Phi = 0 \, ,
\end{equation}
where $\Delta$ denotes the Laplacian on the $(D-2)-$sphere. We assume a radial expansion of the form
\begin{equation} \label{generic Bondi radial expansion of the scalar}
    \Phi(u,r,x^i) = \sum_{n\ge0} r^{-n} \left( \phi^{(n)}(u,x^i) + \log r \, \widetilde{\phi}^{(n)}(u,x^i) \right) ,
\end{equation}
which turns the above equations of motion into the following recursive relation:
\begin{equation}
    \begin{split}
    &\left[ \Delta + n (n-D+3) \right] \phi^{(n)} + \left( 2n-D+4 \right) \partial_u \phi^{(n+1)} + \frac{1}{\ell^2} \left( n+2 \right) \left( n-D+3 \right) \phi^{(n+2)}\\
    &= \left( 2n - D + 3 \right) \widetilde{\phi}^{(n)}+ 2 \, \partial_u \widetilde{\phi}^{(n+1)} + \frac{1}{\ell^2} \left( 2n-D+5 \right) \widetilde{\phi}^{(n+2)} \, ,
    \end{split}
\end{equation}
and similarly for the log terms. We obtain that $\phi^{(0)}$ and $\phi^{(D-1)}$ are undetermined.  
In terms of $\mathcal{F}_\mu := \nabla_\mu \Phi$, the equation of motion takes the form
\begin{equation}
    \frac{1}{r^2} D \cdot \mathcal{F} + \frac{D-2}{r} \left( \mathcal{F}_r - \mathcal{F}_u \right) - \partial_{(u} \mathcal{F}_{r)} + \partial_r \mathcal{F}_r + \frac{1}{\ell^2} \left( r^2 \partial_r^2 + D \, r \right) \mathcal{F}_r = 0 \, ,
\end{equation}
where the round brackets denote  symmetrization on the corresponding indices, $\partial_{(u} \mathcal{F}_{r)} = \partial_{u} \mathcal{F}_{r} - \partial_{r} \mathcal{F}_{u}$, and where $``\cdot"$ stands for contraction of indices on the sphere, raised and lowered using $\gamma^{ij}$ and $\gamma_{ij}$.  

\paragraph{Symplectic renormalization}

In Bondi coordinates the presymplectic potential has radial divergences when $r \to \infty$, so that one has to renormalize the symplectic structure along the lines of section~\ref{sec: Spin-0 FG} and appendix~\ref{app:CPS}. The components of the presymplectic potential \eqref{presymplectic potential massless scalar field strength} and the Lagrangian are 
\begin{align}
        \Theta^r &= r^{D-2} \sqrt{-\gamma} \, \widetilde{\Theta}^r = - r^{D-2} \sqrt{-\gamma} \left( \mathcal{F}_r - \mathcal{F}_u + \frac{r^2}{\ell^2} \, \mathcal{F}_r \right) \delta \Phi \, ,\label{radial thetatilde scalar}\\[1ex]
        \Theta^u &= r^{D-2} \sqrt{-\gamma} \, \widetilde{\Theta}^u = r^{D-2} \sqrt{-\gamma} \, \mathcal{F}_r \, \delta \Phi \, ,\\[1ex]
        \Theta^i &= r^{D-2} \sqrt{-\gamma} \, \widetilde{\Theta}^i = - r^{D-4} \sqrt{-\gamma} \, \mathcal{F}^i \, \delta \Phi \, ,
\end{align}
and 
\begin{equation}
\mathscr{L} = r^{D-2} \sqrt{-\gamma} \, \widetilde{\mathscr{L}} = \frac{\sqrt{-\gamma}}{2} \, r^{D-2} \left[ 2 \, \mathcal{F}_u \, \mathcal{F}_r + \left( 1 + \frac{r^2}{\ell^2} \right) \left( \mathcal{F}_r \right)^2 + \frac{1}{r^2} \, \mathcal{F}^i \, \mathcal{F}_i \right] .
\end{equation}
Upon substituting in \eqref{Spin-0 FG Presymp Radial Eq To take first} one obtains
\begin{equation} \label{Spin-0 Bondi Presymp Radial Eq}
\frac{1}{r} \left(r \, \partial_r + D - 2 \right) \widetilde{\Theta}^r \approx  \delta \widetilde{\mathscr{L}} - \partial_u \widetilde{\Theta}^u - \partial_i \widetilde{\Theta}^i \, .
\end{equation}
which, in its turn, when taking into account the radial expansions 
\begin{equation}
    \widetilde{\Theta}^\mu = \sum_n r^{-n} \left( \widetilde{\Theta}^{\mu}_{(n)} + \log r \, \widetilde{\theta}^{\mu}_{(n)} \right) , \qquad \widetilde{\mathscr{L}} = \sum_n r^{-n} \left( \widetilde{\mathscr{L}}^{(n)} + \log r \, \widetilde{\ell}^{(n)} \right) ,
\end{equation}
yields
\begin{equation}  \label{Spin-0 Bondi Presymp Radial Eq order}
    (D-n-2) \, \widetilde{\Theta}^{r}_{(n)} + \widetilde{\theta}^{r}_{(n)} \approx \delta \widetilde{\mathscr{L}}^{(n+1)} - \partial_u \widetilde{\Theta}^{u}_{(n+1)} - \partial_i \widetilde{\Theta}^{i}_{(n+1)} \, .
\end{equation}
In the following, we shall illustrate along these lines two specific examples in even and odd dimensions, $D=4,5$. 

Let us first consider $D=4$. First, we compute the asymptotic solution space and obtain that the orders $\phi^{(0)}$, $\phi^{(3)}$ are arbitrary functions of the boundary coordinates functions of $u$ and $x^i$. 
Next, we inject the corresponding expansions into the radial component of the presymplectic potential:
\begin{equation}
    \Theta^r = r \sqrt{-\gamma} \, \widetilde{\Theta}^{r}_{(1)} + \sqrt{-\gamma} \, \widetilde{\Theta}^{r}_{(2)} + \mathcal{O}\left(r^{-1}\right) ,
\end{equation}
where
\begin{subequations}
\begin{align}
    \widetilde{\Theta}^{r}_{(1)} &= \delta \phi^{(0)} \left( D \cdot F^{(0)} - \ell^2 \, \partial_u F_u^{(0)} \right) ,\\[1ex]
    \begin{split}
    \widetilde{\Theta}^{r}_{(2)} &= - \frac{1}{2 \ell^2} \Big[ 2 \, F_r^{(4)} \, \delta \phi^{(0)} + \ell^4 \, \Big( 2 \, F_u^{(0)} \, \delta \phi^{(0)} - \partial_u D \cdot F^{(0)} + 2 \, \big( D \cdot F^{(0)}\\
    &\quad - \ell^2 \, \partial_u F_u^{(0)} \big) \, \partial_u \delta \phi^{(0)} \Big) \Big] \, .
    \end{split}
\end{align}
\end{subequations}
So we have a divergent order, $\widetilde{\Theta}^{r}_{(1)}$ to be renormalized. Using the radial renormalization equation \eqref{Spin-0 Bondi Presymp Radial Eq order} we have
\begin{equation}
    \widetilde{\Theta}^{r}_{(1)} = \delta \widetilde{\mathscr{L}}^{(2)} - \partial_u \widetilde{\Theta}^{u}_{(2)} - \partial_i \widetilde{\Theta}^{i}_{(2)} \, ,
\end{equation}
so that
\begin{equation}
    \widetilde{\Theta}^{u}_{(2)} = \ell^2 \, F_u^{(0)} \, \delta \phi^{(0)} \, , \; \widetilde{\Theta}^{i}_{(2)} = - F^i_{(0)} \, \delta \phi^{(0)} \, , \; \widetilde{\mathscr{L}}^{(2)} = - \frac{1}{2} \left( F^i_{(0)} \, F_i^{(0)} - \ell^2 \left( F_u^{(0)} \right)^2 \right) .
\end{equation}
Thus, the counterterm to be added to the presymplectic potential is
\begin{equation}
    \Theta^r_{\text{ct}} = \delta B_{\text{ct}} - \partial_u C^u_{\text{ct}} - \partial_i C^i_{\text{ct}} \, ,
\end{equation}
where
\begin{equation}
    B_{\text{ct}} = - \sqrt{-\gamma} \, \widetilde{\mathscr{L}}^{(2)} \, , \qquad C^u_{\text{ct}} = - \sqrt{-\gamma} \, \widetilde{\Theta}^{u}_{(2)} \, , \qquad C^i_{\text{ct}} = - \sqrt{-\gamma} \, \widetilde{\Theta}^{i}_{(2)} \, ,
\end{equation}
which implies that the renormalized potential takes the following form:
\begin{equation}
    \begin{split}
        \Theta^r_{\text{ren}} &= \lim_{r \to \infty} \left( \Theta^r + \Theta^r_{\text{ct}} \right) = \sqrt{-\gamma} \, \widetilde{\Theta}^r_{(2)}\\
        &= - \sqrt{-\gamma} \bigg[ \frac{1}{\ell^2}\, F_r^{(4)} \, \delta \phi^{(0)} + \frac{\ell^2}{2} \, \Big( 2 \, F_u^{(0)} \, \delta \phi^{(0)} - \partial_u D \cdot F^{(0)} + 2 \, \big( D \cdot F^{(0)}\\
        &\quad - \ell^2 \, \partial_u F_u^{(0)} \big) \, \partial_u \delta \phi^{(0)} \Big) \bigg] \, .
    \end{split}
\end{equation}

In the odd dimensional example of $D=5$, the asymptotic solution space is given by 
    \begin{align}
        \Phi &= \phi^{(0)} + \tfrac{1}{r} \phi^{(1)} + \tfrac{1}{r^2} \phi^{(2)} + \tfrac{1}{r^3} \phi^{(3)}+ \tfrac{1}{r^4}\log r \tilde \phi^{(4)}+ \tfrac{1}{r^4} \phi^{(4)} + \cdots
    \end{align}
where $\phi^{(0)}$ and $\phi^{(4)}$ are arbitrary functions of $u$ and $x^i$. It leads to the following radial expansion of the presymplectic potential
\begin{equation}
    \Theta^r = r^2 \widetilde{\Theta}^r_{(1)} + r \widetilde{\Theta}^r_{(2)} + \log r \, \widetilde{\theta}^r_{(3)} + \widetilde{\Theta}^r_{(3)} + \mathcal{O}(r^{-1}) \, ,
\end{equation}
where the first three terms diverge with $r$. If we add the following counterterm to the presymplectic potential
\begin{equation}
    \Theta^r_{\text{ct}} = \sum_{p=2}^4 \frac{\sqrt{-\gamma}}{p-1} \left( \delta \widetilde{\mathcal{L}}^{(p)} - \partial_u \widetilde{\Theta}^u_{(p)} - \partial_i \widetilde{\Theta}^i_{(p)} \right) ,
\end{equation}
where
\begin{equation}
    - 2\ell^2\widetilde{\mathcal{L}}^{(p)} = \sum_{q=0}^{p-2} \left[ F_r^{(q+2)} F_r^{(p-q)} + \ell^2 \left( F^i_{(q)} F_i^{(p-2-q)} - 2 F_r^{(p-q)} F_u^{(q)} \right) \right] + \ell^2 \sum_{q=0}^{p-4} F_r^{(q+2)} F_r^{(\tfrac{p}{2}-q)} \, ,
\end{equation}
and
\begin{equation}
    \widetilde{\Theta}^u_{(p)} = \sum_{q=0}^{p-2} F_r^{(p-q)} \delta \phi^{(q)} \, , \qquad \widetilde{\Theta}^i_{(p)} = - \sum_{q=0}^{p-2} F_i^{(p-2-q)} \delta \phi^{(q)} \, ,
\end{equation}
one obtains the renormalized potential, whose explicit form is
\begin{equation}
    \begin{split}
        \Theta^r_{\text{ren}} &= \frac{\sqrt{-\gamma}}{24 \ell^2} \bigg\{ - 24 F_r^{(5)} \delta \phi^{(0)} + \ell^4 \bigg[ 2 \delta \phi^{(0)} \bigg( 6 D \cdot F^{(0)} + \ell^2 \partial_u \Big( \ell^2 \partial_u^2 F_u^{(0)} - 3 \partial_u D \cdot F^{(0)}\\
        &\quad + 10 F_u^{(0)} \Big) \bigg) - 3 \ell^4 \Big( \partial_u^2 \delta \phi^{(0)} + 4 \partial_u^2 F_u^{(0)} \partial_u \delta \phi^{(0)} \Big) + 3 D \cdot F^{(0)} \Box \delta \phi^{(0)}\\
        &\quad + 3 \ell^2 \Big( D \cdot F^{(0)} \partial_u^2 \delta \phi^{(0)} + 4 \partial_u D \cdot F^{(0)} \partial_u \delta \phi^{(0)} - \partial_u F_u^{(0)} \Box \delta \phi^{(0)} \Big) \bigg] \bigg\} \, .
    \end{split}
\end{equation}
%

\section{Spin-one fields in Poincar\'e coordinates}
\label{sec:Spin1Poincare}

In this section we compute and renormalize asymptotic charges of massless spin-one fields in AdS space of arbitrary dimension, in Poincar\'e coordinates \ref{app:Poincare}.

\subsection{Solution space}
Maxwell's equations 
$
	\partial_\mu (\sqrt{-g}\,\mathcal F^{\mu\nu})=0\,
$ in Poincar\'e coordinates read
\begin{equation}\label{EOM1Poin}
	\partial^a \mathcal F_{az}=0\,, \qquad
	\frac{1}{z \ell^2}\left( z\partial_z - D + 4\right)\mathcal F_{za} + \partial^b\mathcal F_{ba}=0\, ,
\end{equation}
and combining them with the Bianchi identities, 
$\partial_\mu  \mathcal F_{\nu\rho} + 	\partial_\nu \mathcal F_{\rho\mu}+	\partial_\rho \mathcal  F_{\mu\nu}=0\,$, one obtains
\begin{equation}\label{eqforFab}
	\partial_z\left(
	z^{4-D}\partial_z \mathcal F_{ab}
	\right)
	=
	-z^{4-D} \ell^2 \Box \mathcal F_{ab}\,.
\end{equation}
To begin with, let us explore the solution space in terms of the field strength.

For even $D$ we assume an expansion in $z$ of the schematic form 
\begin{equation}
 \mathcal F (z,x)= \sum_n z^n F^{(n)}(x)\, 
\end{equation}
and get
\begin{equation}\label{eomn1}
	\partial^a F^{(n)}_{az}=0\,,\qquad
	\frac{1}{\ell^2}(n-D+4) F^{(n)}_{za}+\partial^{b}F_{ba}^{(n-1)}=0\, ,
\end{equation}
and
\begin{equation}\label{rrFab}
	\frac{1}{\ell^2}\,n(D-n-3) F_{ab}^{(n)} = \Box F_{ab}^{(n-2)}\,,
\end{equation}
where the first of \eqref{eomn1} is the divergence of the second one, for $n \neq D-4$. 

One sees that \eqref{rrFab} fixes $F_{ab}^{(n)}$ in terms of $F_{ab}^{(0)}$ for any positive even integer $n$,
\begin{equation}\label{}
	F^{(n)}_{ab} = \Bigg(\prod_{q=1}^{\frac{n}{2}}\frac{1}{(D-n-5+2q)}\Bigg)\frac{(\ell^2\Box)^{\frac{n}{2}}}{n!!}\,F_{ab}^{(0)} \, .
\end{equation} 
Similarly, one can fix $F_{ab}^{(D-3+n)}$ for positive even $n$ in terms of $F_{ab}^{(D-3)}$.
Thus, for even $D$, we assume the following structure of the solution as an expansion in $z$,
\begin{subequations}\label{falloff}
	\begin{align}
\mathcal F_{za} & = \sum_{n \geq 0} z^{2n+1} F^{(2n+1)}_{za} + \sum_{n \geq 0} z^{D-4+2n}F_{za}^{(D-4+2n)}\,,\\
\mathcal F_{ab} & = \sum_{n \geq 0} z^{2n} F^{(2n)}_{ab} + \sum_{n \geq 0} z^{D-3+2n}F_{ab}^{(D-3+2n)}\,.
	\end{align}
\end{subequations}
The free data in this case are encoded in $F_{ab}^{(0)}$, which is an arbitrary antisymmetric tensor, and in $F_{az}^{(D-4)}$, which  by the first equation in \eqref{eomn1} is  divergence-free. These two characteristic orders at which the new boundary data appear correspond to the ``source'' and to the ``VEV'' according to the standard terminology.

For odd $D>4$, we include logarithmic terms and expand the field strength according to
\begin{equation}
    \mathcal F (z,x)= \sum_n z^n F^{(n)}(x) + \log z\, \sum_n z^n \widetilde F^{(n)}(x)\,,
\end{equation} 
which modifies \eqref{EOM1Poin} into
\begin{equation}\label{}
		\partial^a F^{(n)}_{az}
  =
		\partial^a \widetilde F^{(n)}_{az}=0\,,\qquad
	\frac{1}{\ell^2}(n-D+4) \widetilde F^{(n)}_{za}+\partial^{b} \widetilde F_{ba}^{(n-1)}=0\,,
\end{equation}
and 
\begin{equation}\label{eomnodd}
	\frac{1}{\ell^2}(n-D+4) F^{(n)}_{za}
	+\frac{1}{\ell^2} \widetilde F^{(n)}_{za}
	+\partial^{b}F_{ba}^{(n-1)}
	=0\, ,
\end{equation}
so as to lead, together with the Bianchi identities, to the  recursion relations
\begin{subequations}
\begin{align}\label{}
0 &= \frac{1}{\ell^2}(n-D+4)(n+1) \widetilde F_{ab}^{(n+1)} + \Box \widetilde F_{ab}^{(n-1)} \, ,\\[5pt]
\label{eomFFtildeRR}
0 &= \frac{1}{\ell^2}(n-D+4)(n+1) F_{ab}^{(n+1)}  + \frac{1}{\ell^2}(2n-D+5)\widetilde F_{ab}^{(n+1)}+ \Box F_{ab}^{(n-1)} \,.
\end{align}
\end{subequations}
In this case, we assume 
\begin{subequations}\label{falloffoddD}
	\begin{align}
		\mathcal F_{za} & = \sum_{n \geq 0} z^{2n+1} F^{(2n+1)}_{za} +
		\log z \, \left( z^{D-4}\widetilde F_{za}^{(D-4)}+\cdots\right) ,\\
		\mathcal F_{ab} & = \sum_{n \geq 0} z^{2n} F^{(2n)}_{ab} +
		\log z \left(z^{D-3}\widetilde F_{ab}^{(D-3)}+\cdots\right) .
	\end{align}
\end{subequations}
It turns out that the logarithmic branch can be fixed in terms of the free data specified by the source, in such a way that $F_{ab}^{(0)}$ remains unconstrained. 

To summarize, the structure of the solution space for arbitrary $D$ is as follows: the arbitrary functions are $F_{ab}^{(0)}$ and $F_{za}^{(D-4)}$, the last one being divergence-free, while 
\begin{subequations}
    \begin{alignat}{5}
        &F_{ab}^{(2n)} = \frac{(-4)^{-n} \ell^{2n} \Gamma\left(\frac{5-D}{2}\right) \Box^{n} F_{ab}^{(0)}}{\Gamma\left(n+1\right) \Gamma\left(\frac{2n+5-D}{2}\right)} \, , \qquad &&F_{ab}^{(2n+1)} = 0 \, , && \hspace{-40pt} 0 < n < \frac{D-3}{2} \, \\
        &F_{za}^{(2n+1)} = \frac{(-4)^{-n} \ell^{2n+2} \Gamma \left( \frac{7-D}{2} \right) \Box^n \partial^b F_{ab}^{(0)}}{(D-5) \Gamma \left( n + 1 \right) \Gamma \left( \frac{2n - D + 7}{2} \right) } \,, \qquad &&F_{za}^{(2n)} = 0 \, ,  && \hspace{-40pt} 0 \leq n < \frac{D-5}{2} \,, \\
        &\widetilde{F}_{ab}^{(D-3)} = - \frac{{\ell^2} \Box}{D-3} F_{ab}^{(D-5)} \, , &&\widetilde{F}_{za}^{(D-4)} = - {\ell^2} \partial^b F_{ab}^{(D-5)} \, .
    \end{alignat}
\end{subequations}
Note that in particular the order $F_{ab}^{(D-3)}$ is fixed in terms of $F_{az}^{(D-4)}$ by
\begin{equation}
    F_{ab}^{(D-3)} = \frac{1}{D-3} \left( \partial_b F_{az}^{(D-4)} - \partial_a F_{bz}^{(D-4)} \right) .
\end{equation}
The falloffs \eqref{falloff} and \eqref{falloffoddD} for even and odd $D$, respectively, capture the two expected branches of solutions associated to radiation (or source) and to static (or VEV) contributions.

Let us now discuss the solution space in terms of the gauge potential $\mathcal{A}_{\mu}$. Upon imposing the Lorenz gauge $\nabla \cdot \mathcal{A} = 0$ and assuming the expansion
\begin{equation} 
    \mathcal{A}(z,x) = \sum_{n \geq 0} z^n A^{(n)}(x) + \log z \sum_{n \geq 0} z^n \widetilde{A}^{(n)}(x) \, ,
\end{equation}
the relations \eqref{eomn1} and \eqref{eomnodd} yield
\begin{subequations}
    \begin{align}
        0 &= (n-1)(n-D+2) A_z^{(n)} + \ell^2 \Box A_z^{(n-2)} + (2n-D+1) \widetilde{A}_z^{(n)} \, ,\\[5pt]
        0 &= n(n-D+3) A_a^{(n)} + \ell^2 \Box A_a^{(n-2)} - 2 \ell^2 \partial_a A_z^{(n-1)} + (2n-D+3) \widetilde{A}_a^{(n)} \, ,
    \end{align}
\end{subequations}
while the gauge condition reads
\begin{equation}
    0 = (n-D+2) A_z^{(n)} + \ell^2 \partial \cdot A^{(n-1)} + \widetilde{A}_z^{(n)} \, .
\end{equation}
The equations determining the log terms are obtained similarly. The solutions, for any $D \ge 4$, are
\begin{subequations}
    \begin{alignat}{5}
        &A_z^{(2n+1)} = \frac{4^{-n} \ell^{2n+2} \Gamma\left( \frac{D-2n-3}{2} \right) \Box^n \partial \cdot A^{(0)}}{(D-3)^2 \Gamma \left(\frac{D-3}{2}\right) \Gamma\left(1+n\right)} \, , \quad &&0 \leq n < \frac{D-2}{2} \,, \\
        &A_a^{(2n)} = \frac{\ell^{2n} \Gamma \left( \frac{D-2n-3}{2} \right) \Box^{n-1} \left( (D-3) \Box A_a^{(0)} - 2 \partial_a \partial \cdot A^{(0)} \right)}{2^{2n+1} \Gamma \left( n + 1 \right) \Gamma \left( \frac{D-1}{2} \right) } \, , \quad && 0 < n < \frac{D-3}{2} \,,
    \end{alignat}
\end{subequations}
while $A_z^{(2n)} = 0$, $A_a^{(2n+1)} = 0$ and
\begin{equation}
    \widetilde{A}_z^{(D-2)} = - \ell^2 \partial \cdot A^{(D-3)} = - \frac{\ell^2 \Box A_z^{(D-4)}}{D-3} \, , \qquad \widetilde{A}_a^{(D-3)} = - \frac{\ell^2}{D-3} \left( \Box A_a^{(D-5)} - 2 \partial_a A_z^{(D-4)} \right) .
\end{equation}
The arbitrary functions of the boundary coordinates are $A_a^{(0)}$, $A_z^{(D-2)}$ and the transverse part of $A_a^{(D-3)}$  (while   $\partial \cdot A^{(D-3)} = \frac{\Box A_z^{(D-4)}}{D-3}$),  and the corresponding radial expansions are given by 
\begin{align}
    \text{even }D\ge4 : \quad
    \begin{split}
    \mathcal{A}_z &= \sum_{n \geq 0} z^{2n+1} A_z^{(2n+1)} + \sum_{n \geq 0} z^{D-2+2n} A_z^{(D-2+2n)} \, ,\\
    \mathcal{A}_a &= \sum_{n \geq 0} z^{2n} A_a^{(2n)} + \sum_{n \geq 0} z^{D-3+2n} A_a^{(D-3+2n)} \, ,
    \end{split}\\[2ex]
    \text{odd }D>4 : \quad
    \begin{split}
    \mathcal{A}_z &= \sum_{n \geq 0} z^{2n+1} A_z^{(2n+1)} + \log z \, z^{D-2} \left( \widetilde{A}_z^{(D-2)} + \dots \right) ,\\
    \mathcal{A}_a &= \sum_{n \geq 0} z^{2n} A_a^{(2n)} +  \log z \, z^{D-3} \left(\widetilde{A}_a^{(D-3)} + \dots \right) .
    \end{split}
\end{align}
In Lorenz gauge the parameter in $\delta_\lambda \mathcal{A}_\mu = \nabla_\mu \lambda$ is constrained by $\nabla^\mu \nabla_{\!\mu} \lambda = 0$. Thus, one can perform on $\lambda$ the same analysis illustrated 
for scalar fields in section~\ref{sec: Spin-0 FG} and find the corresponding solutions \eqref{solution of spin-0 massless field}. In particular, one can use the available free function $\lambda^{(D-1)}$ in the residual gauge parameter to gauge-fix $A_z^{(D-2)}$ to zero. As we shall see below, see e.g.~Eq.~\eqref{ChargeExplEvenP}, $\lambda^{(D-1)}$ does not appear in the boundary charges (in contrast with $\lambda^{(0)}$), so that the transformation needed to achieve this gauge fixing is a small gauge transformation. This shows that the nontrivial boundary data consists of source, i.e.~the free vector $A_a^{(0)}$, and VEV, i.e.~the divergence-free part of $A_a^{(D-3)}$.

As an alternative, in the radial gauge $\mathcal{A}_z = 0$, \eqref{eomn1} and \eqref{eomnodd} become
\begin{subequations}
    \begin{align}\label{eqsaRAD}
        0 &= n \, \partial \cdot A^{(n)} + \partial \cdot \widetilde{A}^{(n)} \, ,\\[5pt]
        0 &= n (D-3-n) A_a^{(n)} + (D-2n-3) \widetilde{A}_a^{(n)} - \ell^2 \Box A_a^{(n-2)} + \ell^2 \partial^b \partial_a A_b^{(n-2)} \, ,
    \end{align}
\end{subequations}
whose solutions are, for $D\ge4$,
\begin{subequations}
    \begin{align}
        &A_a^{(2n)} = \frac{\ell^{2n} \Gamma\left(\frac{5-D}{2}\right) \Box^{n-1} \left( \Box A_a^{(0)} - \partial_a \partial \cdot A^{(0)} \right)}{(-4)^{n}  \Gamma\left(\frac{2n-D+5}{2}\right) \Gamma\left(n+1\right)} \, , \qquad A_a^{(2n+1)} = 0 \, , \qquad 0 < n < \frac{D-3}{2} \,, \\
        \label{eqsAtildeRAD}
        &\widetilde{A}_a^{(D-3)} = \frac{\partial_a \partial \cdot A^{(D-5)} - \Box A_a^{(D-5)}}{D-3} \, ,
    \end{align}
\end{subequations}
with the extra constraints
\begin{equation}\label{eqsAtildeRAD2}
    \partial \cdot A^{(n)} = 0 \, , \qquad  \partial \cdot \widetilde{A}^{(D-3)}= 0 \, ,
\end{equation}
for $n \neq 0, D-3$. Note also that 
\begin{equation}
\partial^a A_a^{(D-3)}=0
\end{equation}
as a consequence of \eqref{eqsaRAD} and \eqref{eqsAtildeRAD2}. Thus, for even dimension
\begin{equation}
    \mathcal{A}_a = \sum_{n \geq 0} z^{2n}  A_a^{(2n)} + \sum_{n \geq 0} z^{D-3+2n} A_a^{(D-3+2n)} \, ,
\end{equation}
while for odd dimension 
\begin{equation}
    \mathcal{A}_a = \sum_{n \geq 0} z^{2n} A_a^{(2n)} + \log z \, z^{D-3} \left( \widetilde{A}_a^{(D-3)} + \ldots \right) .
\end{equation}
The radial orders $A_a^{(0)}$ and the divergence-free part of $A_a^{(D-3)}$ are unconstrained by the above equations of motion. 

\subsection{Holographic renormalization}

The starting point is the regularized action
\begin{equation}\label{}
	S_\mathrm{reg} = -\frac14\int_{z>\epsilon} \text{d}^{D}x \, \sqrt{-g} \mathcal F_{\mu\nu}\mathcal F^{\mu\nu}\,.
\end{equation}
On-shell $\partial_\mu \mathcal A_\nu\,\sqrt{-g}\mathcal F^{\mu\nu}\approx \partial_\mu (\mathcal A_\nu\,\sqrt{-g}\mathcal F^{\mu\nu})$ and thus
\begin{equation}\label{Sreg}
	S_{\mathrm{reg}} \approx 	
	\frac{1}{2\ell\epsilon^{D-4}}\int_{z=\epsilon}\text{d}^{D-1}x\, \mathcal A^{a}\mathcal F_{za}\,.
\end{equation}
Upon expanding the fields for small $z=\epsilon$, we see that this regulated action diverges for $D\ge5$ and one can isolate the divergent terms to then identify the counterterms needed to obtain the renormalized action. Once the latter is available, from its on-shell variation  it is possible to compute the renormalized asymptotic charges.
Note instead that \eqref{Sreg} is finite for $D=4$ thanks to the falloffs \eqref{falloff}.

\subsubsection{Even dimensions}
In order to renormalize \eqref{Sreg}, we shall follow a strategy analogous to the one detailed for the scalar field below Eq.~\eqref{SregulatedScalar214Bas}. Starting from  
\begin{equation}
    S_{\mathrm{reg}} \approx 	
	\int_{z=\epsilon}\text{d}^{D-1}x\, \mathcal A^{a} \mathcal B_{a}\,,\qquad  \mathcal B_{a} = \frac{1}{2\ell\epsilon^{D-4}}\mathcal F_{za}\,,
\end{equation}
in order to define an appropriate subtracted action, it is enough to find a counterterm action $S_{\mathrm{ct}}$ such that
\begin{equation}
     S_{\mathrm{sub}} = S_{\mathrm{reg}} + S_{\mathrm{ct}} 
     \approx 	
	\int_{z=\epsilon}\text{d}^{D-1}x\, \mathcal A^{a} \tilde{\mathcal{B}}_{a}\,,
\end{equation}
where $\tilde{\mathcal{B}}_a$ is finite as $\epsilon\to0$, since $\mathcal A^{a}$ is itself finite in that limit.
To this end, upon inserting the first of \eqref{falloff} in \eqref{Sreg} we have 
\begin{equation}\label{}
	S_\text{reg}
	\approx
	\frac{1}{2\ell\epsilon^{D-4}}
	\int \text{d}^{D-1}x\,
	\mathcal A^a \sum_{n=1}^{D-5} \epsilon^n F^{(n)}_{za} + {\cal O} (1) \, ,
\end{equation}
where the sum runs formally over all $n$ from $1$ to $D-5$ but only the odd terms are nonvanishing. 
Note that, in the present renormalization scheme, we conveniently choose not to expand $\mathcal A^a = \mathcal O(z^0)$.
Similarly to the discussion for the scalar case below Eq.~\eqref{SregulatedScalar214}, this is because we do not aim to obtain the explicit expansion of the regulated action in terms of asymptotic field components $A_\mu^{(k)}$, but rather only its expression in terms of bulk fields, which will suffice in order to obtain a finite presymplectic potential. Let us also stress that this choice cannot affect the bulk-covariant form of the counterterms, except possibly for scheme-dependent terms in the finite piece (and only for odd $D$, as for the scalar field).
This choice is particularly convenient because it allows us to only use the equations of motion for the  field strength; below we will compare this with a different procedure, checking that a different choice only introduces finite local functionals of the source field. 
Using the equations of motion \eqref{eomn1} 
together with  
\begin{equation}\label{subtr0}
	F_{ab}^{(0)} = \mathcal F_{ab}-\sum_{n=2}^\infty \epsilon^n F_{ab}^{(n)}
\end{equation}
near the boundary, we rewrite the regulated action in the form
\begin{equation}\label{}
	S_\text{reg} \approx
 \frac{\ell}{2}
	\int \text{d}^{D-1}x\,
	\frac{\mathcal A^a\partial^b\mathcal F_{ba}}{(D-5)\epsilon^{D-5}}
	+
 \frac{\ell}{2}
	\int \text{d}^{D-1}x\,
	\mathcal A^a \sum_{n=2}^{D-6} 
	\left[
	\frac{1}{D-5-n}-\frac{1}{D-5}
	\right]
	 \frac{\partial^bF_{ba}^{(n)}}{\epsilon^{D-5-n}}\,,
\end{equation}
where the first term determines the most singular counterterm.
Let us introduce the notation $S_{\text{ct}}^{(n)}$ for the $n$th counterterm and $S^{\le n}$ for the sum of the counterterms up to $n$
(in decreasing order of divergence), for later convenience. So
\begin{equation}\label{}
	S_\text{ct}^{(D-5)}
	=
	S_{\text{ct}}^{\le D-5}
	=
 -
 \frac{\ell}{2}
	\int \text{d}^{D-1}x\,
	\frac{\mathcal A^a\partial^b\mathcal F_{ba}}{(D-5)\epsilon^{D-5}}
\end{equation}
and
\begin{equation}\label{}
	S_\text{reg} + S_\text{ct}^{(D-5)}
	=
 \frac{\ell}{2}
	\int \text{d}^{D-1}x\,
	\mathcal A^a \sum_{n=2}^{D-6} 
	\frac{n}{(D-5)(D-5-n)}
	\frac{\partial^bF_{ba}^{(n)}}{\epsilon^{D-5-n}}\,.
\end{equation}
The idea is to use \eqref{rrFab} in order to write a recursion equation for the counterterms. 
Indeed, inserting \eqref{rrFab} and shifting the sum by 2, one finds
\begin{equation}\label{}
S_\text{reg} + S_\text{ct}^{\le D-5}
	=
	\frac{1}{D-5}
 \frac{\ell}{2}
	\int \text{d}^{D-1}x\,
	\mathcal A^a \sum_{n=0}^{D-8} 
	\frac{\ell^2\Box}{(D-7-n)(D-5-n)}
	\frac{\partial^bF_{ba}^{(n)}}{\epsilon^{D-7-n}}\,,
\end{equation}
where, exploiting again \eqref{subtr0}, one can isolate the second counterterm,
\begin{equation}\label{secondct}
	S_\text{ct}^{(D-7)}
	=
 - \frac{\ell}{2}
		\int \text{d}^{D-1}x\,
	\frac{	\mathcal A^a  \ell^2\Box}{(D-7)(D-5)^2}
	\frac{\partial^b\mathcal F_{ba}}{\epsilon^{D-7}},
\end{equation}
thus arriving at 
\begin{equation}\label{}
	S_\text{reg} + S_\text{ct}^{\le D-7}
	=
	\frac{\ell}{2(D-5)^2(D-7)}
	\int \text{d}^{D-1}x\,
	\mathcal A^a \sum_{n=2}^{D-8} 
	\frac{n(2D-12-n)}{(D-7-n)(D-5-n)}
	\frac{\ell^2\Box\partial^bF_{ba}^{(n)}}{\epsilon^{D-7-n}}\,.
\end{equation}
The general structure is therefore
\begin{equation}\label{Poincare counterterms}
	S_\text{ct}^{(D-5-i)}\\
	=
 -
 \frac{\ell}{2}
	\int \text{d}^{D-1}x\,
	c^{(i)}\,
	\frac{\mathcal A^a(\ell^2\Box)^{i/2}\partial^b\mathcal F_{ba}}{\epsilon^{D-5-i}}
\end{equation}
and
\begin{equation}\label{}
		S_\text{reg}
  +
		S_\text{ct}^{\le D-3-i}\\
	=
 \frac{\ell}{2}
	\int \text{d}^{D-1}x\,
	\sum_{n=0}^{D-6-i}
	c^{(i)}_n\,
	\frac{\mathcal A^a(\ell^2\Box)^{i/2}\partial^b F^{(n)}_{ba}}{\epsilon^{D-5-n}}
\end{equation}
with
\begin{equation}\label{counterterm coefficients}
	c^{(i)}\, = c^{(i)}_{0}\,
	\,\qquad
	c^{(i)}_{n} = \frac{c^{(i-2)}_{n+2}-c^{(i-2)}}{(n+2)(D-5-n)}
 \,,\qquad
 {
 c_n^{(0)} = \frac{1}{D-5-n}
 }\,.
\end{equation}
Upon integrating by parts one finds the general form of the counterterm 
\begin{equation}\label{counterterms1evenD}
    S_{\text{ct}}^{(\epsilon)} = + \frac{{\ell}}{4 (D-5) \epsilon^{D-5}} \int_{z=\epsilon} \text{d}^{D-1} x \, \mathcal{F}_{ab} \left( 1 + \sum_{k=1}^{\frac{D-6}{2}} {\ell^{2k}} \epsilon^{2k} c_N^{(k+1)} \Box^k \right) \mathcal{F}^{ab} \, ,
\end{equation}
up to a boundary-of-the-boundary term, where the coefficients {$c_N^{(k)}=(D-5) c^{2(k-1)}$ (here $N$ is just a label for ``normalized'', not an additional variable) can be found by means of the recursion relation \eqref{counterterm coefficients} or by the equivalent one}
\begin{equation} \label{recursion counterterm}
    c_N^{(k)} = - \frac{ (-4)^{-k+1} (k-1) \Gamma\left(\frac{-D+5}{2}\right) }{ \Gamma\left(k\right) \Gamma\left(\frac{2k-D+5}{2}\right) } - \sum_{q=0}^{k-3} \left( c_N^{(q+2)} \frac{ (-4)^{-k+2+q} \Gamma\left(\frac{-D+5}{2}\right) }{ \Gamma\left({k-1-q}\right) \Gamma\left(\frac{2k-D+1-2q}{2}\right) } \right) ,
\end{equation}
{with}
\begin{equation}
{
c_N^{(1)}=1\,,\qquad c_N^{(2)} =  \frac{1}{(D-5)(D-7)}\,.
}
\end{equation}
Let us note that the expression \eqref{counterterms1evenD} for the counterterms can be rewritten in the following manifestly diffeomorphism invariant way,
\begin{equation}\label{diffinvEvenDs1}
S_{\text{ct}}^{(\epsilon)} = + \frac{\ell}{4 (D-5)} \int_{z=\epsilon}  \text{d}^{D-1} x \, \sqrt{-\gamma} \,
\gamma^{ac}\gamma^{bd}
\mathcal{F}_{ab} \left( 1 + \sum_{k=1}^{\frac{D-6}{2}}  c_N^{(k+1)} \ell^{2k} \Box_\gamma^k \right) \mathcal{F}_{cd}
\end{equation}
where all explicit powers of $\epsilon$ have been reabsorbed into the induced metric $\gamma_{ab} = \eta_{ab}/\epsilon^2$ on the surface $z=\epsilon$ and into $\Box_\gamma = \gamma^{ab}\partial_a\partial_b$.
There are two possibilities to do away with the boundary-of-the-boundary term. The first one is to assume that we are working with field configurations that fall off in the early past and in the far future at the boundary:
\begin{equation}\label{bbc}
	\mathcal F_{ab} =0\,\ \text{on the boundary of }z=\epsilon\,.
\end{equation}
The second one is to add a corner term to the action that cancels it, whose general form is 
\begin{equation}\label{thecornerterm}
    S_{\text{corner}}^{(\epsilon)} = \frac{{\ell}}{2 (D-5) \epsilon^{D-5}} \int_{z=\epsilon} \text{d}^{D-1} x \, \partial^b \left[ \mathcal{A}^a \left( 1 + \sum_{k=1}^{\frac{D-6}{2}} {\ell^{2k}} \epsilon^{2k} c_N^{(k+1)} \Box^k \right) \mathcal{F}_{ba} \right] .
\end{equation}
Of course this will not interfere with the variational principle, since it only involves derivatives tangential to the boundary. 

Altogether,  the on-shell variation of the subtracted action, $S_{\text{sub}}^{(\epsilon)} = S_{\text{reg}}^{(\epsilon)} + S_{\text{ct}}^{(\epsilon)}$, is given by  
\begin{equation}
\delta S_{\text{sub}}^{(\epsilon)} \approx \int_{z=\epsilon} \text{d}^{D-1} x \,  \delta \mathcal{A}^a \left[ \frac{\mathcal{F}_{za}}{\ell \epsilon^{D-4}} - \frac{{\ell}}{(D-5) \epsilon^{D-5}} \left( 1 + \sum_{k=1}^{\frac{D-6}{2}} {\ell^{2k}} \epsilon^{2k} c_N^{(k+1)} \Box^k \right) \partial^b \mathcal{F}_{ab} \right]
\end{equation}
which, after taking the limit $\epsilon \to 0$, gives us the following variation of the renormalized on-shell action
\begin{align} \label{deltaSrenEvenP}
\delta S_{\text{ren}} = \lim_{\epsilon \to 0} \delta S_{\text{sub}}^{(\epsilon)} \approx \frac{1}{{\ell}} \int \text{d}^{D-1} x \,  \delta \mathcal{A}^a F_{za}^{(D-4)} 
\end{align}
and the boundary conserved current
\begin{equation} \label{CurrentEvenP}
\langle J_a \rangle =  \frac{\delta S_{\text{ren}}}{\delta \mathcal{A}^{a}} = \frac{1}{\ell} F_{za}^{(D-4)} \,.
\end{equation}
In this scheme, where we haven't expanded the Maxwell field and have inverted the expansions, we can provide a covariant expression for the presymplectic potential and symplectic structure in terms of the bulk fields:
\begin{equation}
 \delta S_{\text{sub}}^{(\epsilon)} \approx \int_{z=\epsilon} \text{d}^{D-1} x \, \Theta_{\text{sub}}^{(\epsilon)} \, , \qquad 
 \Omega = \frac{1}{{\ell}} \int_{z=0} \text{d}^{D-1} x \, \delta \mathcal{A}^a \, \delta F_{za}^{(D-4)}\, ,
 \end{equation}
where $F_{za}^{(D-4)} = (D-3) A_a^{(D-3)} - \partial_a A_z^{(D-4)}$ (in both gauges).
In particular, looking at the variation along a gauge parameter $\lambda$, we can rewrite the latter as
\begin{equation}
    \delta_\lambda S_{\text{ren}} = \frac{1}{{\ell}} \int \text{d}^{D-1} x \, \partial^a \left(  \lambda F_{za}^{(D-4)} \right),
\end{equation}
where we used $\partial^a F_{za}^{(D-4)} = 0$ from \eqref{eomn1}. 
Correspondingly, the charge flux across the boundary is given by
\begin{equation}\label{ChargeEvenP}
    \Delta Q^{(\lambda)} = \frac{1}{{\ell}} \int \text{d}^{D-1} x \, \partial^a \left(  \lambda F_{za}^{(D-4)} \right) ,
\end{equation}
and the surface charges read
\begin{equation}\label{ChargeExplEvenP}
    Q^{(\lambda)} = -\frac{1}{{\ell}} \int \text{d}^{D-2} x \, \left(  \lambda F_{z0}^{(D-4)} \right) .
\end{equation}
Let us note that the time derivative $\partial_0 Q^{(\lambda)}$ manifestly vanishes only for the standard electric charge ($\lambda=1$). Let us now discuss in more detail  the counterterm structure in a couple of examples.

In $D=6$ there is only one counterterm which is given by
\begin{equation}
S_{\mathrm{ct}} = -
\frac{1}{2\ell\epsilon}\int_{z=\epsilon}\text{d}^5x\, \mathcal A^{a} \partial^b\mathcal F_{ba} 
\end{equation}
so that, on shell
\begin{equation}\label{splitting}
S_{\mathrm{reg}} \approx
\frac{\ell}{2\epsilon}\int_{z=\epsilon}\text{d}^5x\,  \partial^b (\mathcal A^{a}\mathcal F_{ba}) -
\frac{\ell}{4\epsilon}\int_{z=\epsilon}\text{d}^5x\, \mathcal F_{ab} \mathcal F^{ab} +\mathcal O(1)\,.
\end{equation}
Of the two options to do away with the boundary-of-the-boundary  we simply use \eqref{bbc} to drop the first term in the second line so that the subtracted action is
\begin{equation}\label{renaction16}
	S_{\text{sub}} = \int_{z>\epsilon} \text{d}^6x \left(
	-\frac14\,\sqrt{-g}\,\mathcal F_{\mu\nu}\mathcal F^{\mu\nu}
	\right)
	-
	\frac{\ell}{\epsilon}\int_{z=\epsilon}\text{d}^5x \left( -\frac14\mathcal F_{ab} \mathcal F^{ab}\right);
\end{equation}
note that this expression is exactly gauge invariant.
Its on-shell variation gives 
\begin{equation}\label{}
	\begin{split}
		\delta S_{\text{sub}} &\approx
		\int_{z=\epsilon}\text{d}^5x\,
		\delta \mathcal A^a
		\left(
		\frac{1}{\ell \epsilon^2}\mathcal F_{za}
		-
		\frac{\ell}{\epsilon}
		\partial^b \mathcal F_{ba},
		\right)
	\end{split}
\end{equation}
and expanding for small $\epsilon$, we find 
\begin{equation}\label{lepsepsl}
	\frac{1}{\ell \epsilon^2}\mathcal F_{za}
	-
	\frac{\ell}{\epsilon}
	\partial^b \mathcal F_{ba}
	=
	\frac{\ell}{\epsilon}
	\left(
	\frac{1}{\ell^2}F^{(1)}_{za}
	-
	\partial^b F^{(0)}_{ba}
	\right)
	+
	\frac{1}{\ell}\,F^{(2)}_{za}+\mathcal O(\epsilon)\,,
\end{equation}
where the divergent term vanishes due to \eqref{eomn1}. 
We conclude that
\begin{equation}\label{deltatotheta}
	\delta S_\mathrm{ren} \approx 
	 \frac{1}{\ell}\int_{z=0} \text{d}^5x\,\delta \mathcal A^{a} F^{(2)}_{za}\,.
\end{equation}
We can thus identify the conserved current \eqref{CurrentEvenP} and the surface charge \eqref{ChargeExplEvenP}, for $D=6$.

In $D=8$ a leading and a subleading term need to be taken into account. Indeed, there are two divergent terms 
\begin{equation}\label{}
	S_{\mathrm{reg}} \approx 
	\frac{1}{2\ell\epsilon^{4}}\int_{z=\epsilon}\text{d}^7x\, \mathcal A^{a}\left(\epsilon F_{za}^{(1)}+\epsilon^3 F_{za}^{(3)}\right)+\mathcal O(1)\,.
\end{equation}
Following the steps from \eqref{subtr0} to \eqref{secondct} we obtain
\begin{equation}\label{}
	S_{\mathrm{reg}} \approx 
	\frac{\ell}{2}\int_{z=\epsilon}\text{d}^7x\, \mathcal A^{a} \partial^b\left(\frac{1}{3\epsilon^3} \mathcal F_{ba} + \frac{1}{9\epsilon} \ell^2\Box \mathcal F_{ba}\right)+\mathcal O(1)\,,
\end{equation}
and integrating by parts
\begin{equation}\label{}
\begin{split}
	S_{\mathrm{reg}} &\approx 
\frac{\ell}{2}\int_{z=\epsilon}\text{d}^7x\,\partial^b\left[ \mathcal A^{a} \left(\frac{1}{3\epsilon^3} \mathcal F_{ba} + \frac{1}{9\epsilon} \ell^2\Box \mathcal F_{ba}\right)\right]
\\
& \quad -\frac{\ell}{4}\int_{z=\epsilon}\text{d}^7x\, \mathcal F^{ab} \left(\frac{1}{3\epsilon^3} \mathcal F_{ab} + \frac{1}{9\epsilon} \ell^2\Box \mathcal F_{ab}\right)+\mathcal O(1)\,,
\end{split}
\end{equation}
where the second line  identifies the counterterms. The subtracted action is therefore
\begin{equation}\label{ren8}
	S_{\text{sub}}
	=
	\int_{z>\epsilon} \text{d}^8x \left(
	-\frac14\,\sqrt{-g}\,\mathcal F_{\mu\nu}\mathcal F^{\mu\nu}
	\right)
	-
	\frac{\ell}{4}\int_{z=\epsilon}\text{d}^7x \left( -\frac{1}{3\epsilon^3}\mathcal F_{ab} \mathcal F^{ab} - \frac{1}{9\epsilon} \mathcal F_{ab} \ell^2\Box\mathcal F^{ab}\right).
\end{equation}
In the on-shell variation 
\begin{equation}\label{dSren8}
	\delta S_\text{sub}
	\approx
	\int_{z=\epsilon}
	\text{d}^7x\,
	\delta \mathcal A^a
	\left(
	\frac{1}{\ell\epsilon^4}\mathcal F_{za}
	-
	\frac{\ell}{3\epsilon^3} \partial^b\mathcal F_{ba}
	-\frac{\ell}{9\epsilon} \ell^2\Box\partial^b \mathcal F_{ba}
	\right) ,
\end{equation}
both divergent contributions arising from the expansion of the first term cancel, leaving behind
\begin{equation}\label{}
	\delta S_\text{ren}
	\approx
	\frac{1}{\ell}
	\int_{z=0}
	\text{d}^7x\,
	\delta \mathcal A^a
	 F^{(4)}_{za}\,,
\end{equation}
which matches \eqref{deltaSrenEvenP} for $D=8$, and reproduces current \eqref{CurrentEvenP} and the surface charge \eqref{ChargeExplEvenP}, for $D=8$.

\subsubsection{Odd dimensions}

From \eqref{Sreg} and \eqref{falloffoddD}  we see that in odd $D\ge5$ there will be a single counterterm associated to a logarithmic divergence plus a collection of counterterms associated to the power-like divergences:
\begin{equation}\label{}
	S_\text{reg}
	=
	S_\text{st}
	+
	S_\text{log}
\end{equation}
with
\begin{equation}\label{}
	S_\text{st}
	=
	\frac{1}{2\ell\epsilon^{D-4}}
	\int \text{d}^{D-1}x\,
	\mathcal A^a \sum_{n=1}^{D-6} \epsilon^n F^{(n)}_{za}\,,
	\qquad
	S_\text{log} = - \frac{\ell}{2} \,\log\epsilon \int \text{d}^{D-1}x \mathcal A^a \partial^b F_{ba}^{(D-5)}\,.
\end{equation}
Using the equation of motion \eqref{eomFFtildeRR}, one finds
\begin{equation}\label{}
	S_\text{log} = - \frac{\ell}{2} \,\log\epsilon \int \text{d}^{D-1}x \frac{\mathcal A^a  (\ell^2 \Box)^{\frac{D-5}{2}}}{[(D-5)!!]^2}\,\partial^b \mathcal F_{ba}\,,
\end{equation}
while the structure of $S_\text{st}$ is the same as that of $S_\text{reg}$ in even $D$, i.e.~its counterterms are given by
\eqref{Poincare counterterms} with coefficients \eqref{counterterm coefficients}. The recursion relation starts with
\begin{equation}\label{}
	c^{(0)}_n = \frac1{D-5-n} 
\end{equation}
where of course the $n=D-5$ terms is absent.
Nevertheless, we do have a finite counterterm $S^{(0)}_\text{ct}$ associated to $c^{(D-5)}$.

Equivalently, {for $D>5$ (see \eqref{Sren5} below for $D=5$),} in order to regularize the on-shell action, we can add a counterterm action that differs from the even-dimensional one only by the 
presence of a term in $\log \epsilon$, 
\begin{equation}\label{ctoddDs1}
    S_{\text{ct}}^{(\epsilon)} = \int_{z=\epsilon} \frac{\text{d}^{D-1} x \, \ell}{4 (D-5) \, \epsilon^{D-5}} \mathcal{F}_{ab} \left( 1 + \sum_{k=1}^{\frac{D-7}{2}} (\ell  \epsilon)^{2k} c_N^{(k+1)} \Box^k - (\ell \epsilon)^{D-5} \log \epsilon \, c_L \Box^{\frac{D-5}{2}} \right) \mathcal{F}^{ba} \, ,
\end{equation}
and takes this suitable expression thanks to the further addition of a corner term,
\begin{equation}
S_\text{corner}^{(\epsilon)}
=
    \int_{z=\epsilon} \frac{ \text{d}^{D-1} x\, \ell \, \partial^b}{2 (D-5) \epsilon^{D-5}} \left[ \mathcal{A}^a \left( 1 + \sum_{k=1}^{\frac{D-7}{2}} (\ell \epsilon)^{2k} c_N^{(k+1)} \Box^k - (\ell \epsilon)^{D-5} \log \epsilon \, c_L \, \Box^{\frac{D-5}{2}} \right) \mathcal{F}_{ba} \right] ,
\end{equation}
with the coefficients $c_N^{(k)}$ determined by the recursive relation \eqref{recursion counterterm}, while the coefficient for the logarithmic term is
\begin{equation}
    c_L = 
    {\frac{2^{6-D}}{\Gamma\left(\frac{D-3}{2}\right) \Gamma\left(\frac{D-5}{2}\right)}}.
\end{equation}
Let us also rewrite \eqref{ctoddDs1} by introducing the induced metric $\gamma_{ab}$ on the cutoff surface, in a manner similar to \eqref{diffinvEvenDs1},
\begin{equation}\label{ctoddDs1almostdiff}
    S_{\text{ct}}^{(\epsilon)} = \int_{z=\epsilon} \frac{\ell\, \text{d}^{D-1} x}{4 (D-5)}\,\sqrt{-\gamma} \, \gamma^{ac} \gamma^{bd}\mathcal{F}_{ab}\left( 1 + \sum_{k=1}^{\frac{D-7}{2}} \ell^{2k} c_N^{(k+1)} \Box_\gamma^k - \ell^{D-5} \log \epsilon \, c_L \Box_\gamma^{\frac{D-5}{2}} \right) \mathcal{F}_{cd} \,,
\end{equation}
where the only leftover explicit dependence on $\epsilon$ enters via $\log\epsilon$.
The on-shell variation of the subtracted action yields $\delta S_{\text{sub}}^{(\epsilon)} \approx \int_{z=\epsilon} \Theta_{\text{sub}}^{(\epsilon)} \text{d}^{D-1} x$, where
\begin{equation}
    \Theta_{\text{sub}}^{(\epsilon)} = \frac{\delta \mathcal{A}^a}{\ell \epsilon^{D-5}} \left[ \frac{\mathcal{F}_{za}}{\epsilon} - \frac{{\ell}^2}{(D-5)} \left( 1 + \sum_{k=1}^{\frac{D-7}{2}} {(\ell \epsilon)^{2k}} c_N^{(k+1)} \Box^k - {(\ell \epsilon)^{D-5}} \log \epsilon \, c_L \Box^{\frac{D-5}{2}} \right) \partial^b \mathcal{F}_{ab} \right] .
\end{equation}
In the limit $\epsilon \to 0$, the last equation gives us the following variation of the renormalized on-shell action
\begin{align}
\delta S_{\text{ren}} = \lim_{\epsilon \to 0} \delta S_{\text{sub}}^{(\epsilon)} \approx \frac{1}{{\ell}} \int \text{d}^{D-1} x \,  \, \delta \mathcal{A}^a F_{za}^{(D-4)} \, .
\end{align}
In this case, $F_{za}^{(D-4)} = (D-3) A_a^{(D-3)} + \widetilde{A}_a^{(D-3)} - \partial_a A_z^{(D-4)}$ (in both gauges).
Along a gauge parameter $\lambda$, the latter reduces on-shell to a corner term
\begin{equation}
    \delta_\lambda S_{\text{ren}} = \frac{1}{{\ell}} \int \text{d}^{D-1} x \, \partial^a \left(  \, \lambda F_{za}^{(D-4)} \right) ,
\end{equation}
so that the conserved current and the charge are given by the same expressions as in the even dimensional case, \eqref{CurrentEvenP} and \eqref{ChargeExplEvenP}, respectively.

The case of $D=5$ provides the simplest example to this effect, with the only the divergent term being the logarithmic one
\begin{equation}\label{}
	S_\mathrm{reg} \approx - \frac{\ell}{2}\log\epsilon \int_{z=\epsilon} \text{d}^{4}x\,\mathcal A^a \partial^b\mathcal F_{ba} + \mathcal O(1)\,.
\end{equation}
Assuming that \eqref{bbc} holds, we can integrate by parts this term and write the subtracted action as follows
\begin{equation}\label{Sren5}
	S_\mathrm{ren} = \int_{z>\epsilon} \text{d}^5x\left(
	-\frac14 \sqrt{-g}\,\mathcal F_{\mu\nu}\mathcal F^{\mu\nu}
	\right)
	+
	\ell\log\epsilon\int_{z=\epsilon}\text{d}^4x\left(
	-\frac14 \mathcal F_{ab}\mathcal F^{ab}
	\right).
\end{equation}
Taking the variation on shell and using the condition \eqref{bbc} to drop the boundary of the boundary contribution,
\begin{equation}\label{dSren5}
	\begin{split}
	\delta S_{\text{ren}} 
&\approx
\frac{1}{\ell\epsilon}
\int_{z=\epsilon}\text{d}^4x
\mathcal F_{za}
\delta \mathcal A^a
+
\ell\log\epsilon
\int_{z=\epsilon}\text{d}^4x\,\delta \mathcal A_b \,\partial_a\mathcal F^{ab}\\
&=\int_{z=\epsilon}\text{d}^4x\,\delta \mathcal A^a \left(
\frac{1}{\ell\epsilon}\mathcal F_{za}+\ell\log\epsilon\, \partial^b \mathcal F_{ba}
\right).
	\end{split}
\end{equation}
Expanding the quantity within round parentheses,
\begin{equation}\label{}
	\frac{1}{\ell\epsilon}\mathcal F_{za}+\ell\log\epsilon\, \partial^b \mathcal F_{ba}
	=
	\log\epsilon \left( \frac{1}{\ell} \widetilde F^{(1)}_{za}
	+
	\ell
	\partial^b F_{ba}^{(0)}
	\right)
	+
	\frac{1}{\ell}\,F^{(1)}_{za}+o(1)\,,
\end{equation}
and the coefficient of the singularity vanishes on the equations of motion thanks to \eqref{eomnodd} evaluated for $n=1$.
In conclusion,
\begin{equation}\label{}
\delta S_{\mathrm{ren}} \approx \frac{1}{\ell}\int_{z=0} \text{d}^4x\, \delta \mathcal{A}^a F^{(1)}_{za}
\end{equation}
and therefore
\begin{equation}\label{}
	\langle J_a\rangle = \frac{\delta S_{\mathrm{ren}}}{\delta \mathcal{A}^a}
	=  \frac{1}{\ell}\,F_{za}^{(1)}\,,
\end{equation}
and the equations of motion indeed ensure that this is a conserved current, since $\partial^a F^{(n)}_{za}\approx 0$.
Correspondingly, the charge flux that we get is 
\begin{equation}\label{Q5}
	\Delta Q^{(\lambda)} 
	\approx \frac{1}{\ell} \int_{z=0} \text{d}^4x\, \partial^a\lambda\, F^{(1)}_{za} 
	\approx \frac{1}{\ell} \int_{z=0} \text{d}^4x\, \partial^a (\lambda\, F^{(1)}_{za})\,, 
\end{equation}
where we have used the equations of motion $\partial^a F^{(n)}_{za}\approx 0$. 

The first case in which both a power divergence and a logarithm need to be taken into account simultaneously is the case $D=7$.
Starting again from \eqref{Sreg} for $D=7$,
\begin{equation}\label{}
	S_{\text{reg}} \approx 	\frac{1}{2\ell\epsilon^{3}}\int_{z=\epsilon}\text{d}^6x\, \mathcal A^{a}\mathcal F_{za}
\end{equation}
and substituting the falloffs \eqref{falloffoddD}
 we have
\begin{equation}\label{divergentD7}
	S_{\mathrm{reg}} \approx 
	\frac{1}{2\ell \epsilon^2}\int_{z=\epsilon}\text{d}^6x\, \mathcal A^{a} F^{(1)}_{za}
	+
	\frac{1}{2\ell}\,\log\epsilon\int_{z=\epsilon}\text{d}^6x\, \mathcal A^{a}\widetilde F^{(3)}_{za}+\mathcal O(1)\,.
\end{equation}
The equation of motion \eqref{eomnodd}, setting $n=1$ and $n=3$, gives
\begin{equation}\label{eom7-1}
	F_{za}^{(1)} = \frac{\ell^2}{2}\,\partial^b F_{ba}^{(0)}\,,\qquad
	\widetilde F_{za}^{(3)} = -\ell^2\partial^bF_{ba}^{(2)}\,,
\end{equation}
while the recursion relation \eqref{eomFFtildeRR} yields
\begin{equation}\label{eom7-2}
	F_{ab}^{(2)} = \frac{1}{4}\,\ell^2\Box F_{ab}^{(0)} \implies \widetilde F_{za}^{(3)}=-\frac{\ell^2}{4}\,\ell^2\Box \partial^ bF_{ba}^{(0)}\,.
\end{equation}
Substituting into \eqref{divergentD7} and using $F_{ab}^{(0)}= \mathcal{F}_{ab}-z^2 \,F_{ab}^{(2)}+\cdots$, we find
\begin{equation}\label{}
	S_\mathrm{reg} \approx -\frac{\ell}4 \int_{z=\epsilon} \text{d}^6x\,\mathcal F^{ab} \left(
	\frac1{2\epsilon^2}
	-
	\frac{1}{8}\,\ell^2 \Box
	-
	\frac{1}{4}\,\log\epsilon \ell^2 \Box
	\right) \mathcal F_{ab} + \mathcal O(1)\,,
\end{equation}
after integrating by parts in the usual way.
Note that the second term in the round parentheses gives a finite contribution as $\epsilon\to0$. Since in the minimal subtraction scheme we simply identify all divergent terms of the regularized action, it could be natural to drop it. However, we may also keep it because it actually cancels a finite term arising from the first term when re-expanded in $\epsilon$. 
We are thus led to define the renormalized action as
\begin{equation}\label{Sren7}
	S_\mathrm{ren} = \int_{z>\epsilon} \text{d}^7x\left(
	-\frac14 \sqrt{-g}\,\mathcal F_{\mu\nu}\mathcal F^{\mu\nu}
	\right)
	+
	\frac{\ell}4 \int_{z=\epsilon} \text{d}^{6}x\,\mathcal F^{ab} \left(
	\frac1{2\epsilon^2}
	-
	\frac{1}{8}\,\ell^2\Box
	-
	\frac{1}{4}\,\log\epsilon \ell^2\Box
	\right) \mathcal F_{ab}.
\end{equation}

Taking the variation on shell and using the condition \eqref{bbc} to drop the boundary of the boundary contribution,
\begin{equation}\label{dSren7}
		\delta S_{\text{ren}} 
		\approx
		\int_{z=\epsilon}\text{d}^6x\,
		\delta \mathcal{A}^{a}
		\left[
		\frac{1}{\ell\epsilon^3}\,
		\mathcal F_{za}
	-
	\ell
	\left(
	\frac1{2\epsilon^2}
	-
	\frac{1}{8}\,\ell^2\Box
	-
	\frac{1}{4}\,\log\epsilon \ell^2\Box
	\right) \partial^b\mathcal F_{ba}\right].
\end{equation}
Expanding the quantity within square brackets, we have
\begin{equation}\label{}
	\begin{split}
	\left[\,\cdots\right]_a
	&=
	\frac{1}{\ell\epsilon^2}\left(\mathcal F^{(1)}_{za}-\frac{\ell^2}{2}\partial^bF_{ba}^{(0)}\right)
	+
	\frac{1}{\ell}\,F^{(3)}_{za}
	+
	\frac{\log\epsilon}{\ell}\left(
	\widetilde F_{za}^{(3)}
	+\frac{\ell^2}{4}\,\ell^2\Box \partial^b F_{ba}^{(0)}
	\right)\\
	&\quad +
	\frac{\ell}{2}\,\partial^b\left(\mathcal F^{(2)}_{za}-\frac{1}{4}\,\ell^2\Box \partial^b F_{ba}^{(0)}\right)
	+\mathcal O(1)\,.
	\end{split}
\end{equation}
All round parenthesis in this expression vanish by the equations of motion \eqref{eom7-1}, \eqref{eom7-2},
including the ``spurious'' finite term coming from the variation of the counterterm.
In conclusion,
\begin{equation}\label{}
	\delta S_{\mathrm{ren}} \approx \frac{1}{\ell}\int_{z=0} \text{d}^6x\, \delta \mathcal{A}^a F^{(3)}_{za}
\end{equation}
and therefore
\begin{equation}\label{}
	\langle J_a\rangle = \frac{\delta S_{\mathrm{ren}}}{\delta \mathcal A^a}
	=  \frac{1}{\ell}\,F_{za}^{(3)}\,,
\end{equation}
and the equations of motion indeed ensure that this is a conserved current, since $\partial^a F^{(n)}_{za}\approx 0$.
Correspondingly, the charge flux that we get is 
\begin{equation}\label{Q7}
	\Delta Q^{(\lambda)} 
	\approx \frac{1}{\ell} \int_{z=0} \text{d}^6x\, \partial^a\lambda\, F^{(3)}_{za} 
	\approx \frac{1}{\ell} \int_{z=0} \text{d}^6x\, \partial^a (\lambda\, F^{(3)}_{za})\,, 
\end{equation}
where we have used the equations of motion $\partial^a F^{(n)}_{za}\approx 0$.

\subsection{Symplectic renormalization} \label{sec:symplectic P}

In order to obtain finite charges an alternative procedure is to directly renormalize the symplectic structure of the theory, as we saw in section~\ref{sec: Spin-0 FG} for the scalar case. To this end,  from the Maxwell Lagrangian 
\begin{equation}
    \mathscr{L} = - \frac{\sqrt{-g}}{4} \mathcal{F}^{\mu\nu} \mathcal{F}_{\mu\nu} \, ,
\end{equation}
one derives the presymplectic potential (see appendix~\ref{app:CPS})
\begin{equation} \label{Spin-1 Presymp defining relation}
 \Theta^\mu(\mathcal{A},\delta \mathcal{A}) = - \sqrt{-g} \, \mathcal{F}^{\mu\nu} \delta \mathcal{A}_\nu \, .
\end{equation}
The main idea of the symplectic renormalization is to exploit the ambiguities of the presymplectic potential \eqref{ambiguities def} in order to cancel radial divergences. 
In general, explicitly determining the form of the counterterm ambiguities in \eqref{ambiguities def} can be impractical. However, a systematic way of obtaining such terms has been proposed in \cite{Papadimitriou:2005ii, Freidel:2019ohg}. In this approach, contrary to the holographic renormalization, we need not only the solution space of the field strength but also that of the gauge field and therefore this procedure depends in principle on the gauge. 

Again, since the boundary has the orientation $n_\mu = \delta_\mu^z$ in the Poincar\'e patch, the charges are computed using $\Theta^z$. If we factor out the off-shell radial dependence of $\Theta^{\mu}$
\begin{equation}
    \begin{split}
        \Theta^z &= z^{-(D-4)} \widetilde{\Theta}^z \, , \qquad \widetilde{\Theta}^z = \frac{1}{\ell} \mathcal{F}_{az} \delta \mathcal{A}^a \, ,\\
        \Theta^a &= z^{-(D-4)} \widetilde{\Theta}^a \, , \qquad \widetilde{\Theta}^a = \frac{1}{\ell} {\mathcal{F}_z}^{a} \delta \mathcal{A}_z - \ell \mathcal{F}^{ab} \, \delta \mathcal{A}_b \, ,
    \end{split}
\end{equation}
and similarly for the Lagrangian
\begin{equation}
\mathscr{L} = z^{-(D-4)} \widetilde{\mathscr{L}} \, , \qquad \widetilde{\mathscr{L}} =  \frac{1}{2 \ell} {\mathcal{F}^a}_{z} \mathcal{F}_{za} - \frac{\ell}{4} \mathcal{F}^{ab} \mathcal{F}_{ab} \, ,
\end{equation}
we obtain the asymptotic renormalization equation 
\begin{equation} \label{Spin-1 Poincar'e Presymp Radial Eq}
\frac{1}{z} \left(z \partial_z - (D-4) \right) \widetilde{\Theta}^z \approx \delta \widetilde{\mathscr{L}} - \partial_a \widetilde{\Theta}^a \, .
\end{equation}
Upon further expanding radially $ \widetilde{\Theta}^\mu$ and $\widetilde{\mathscr{L}}$,
\begin{equation}
    \widetilde{\Theta}^\mu = \sum_n z^n \left( \widetilde{\Theta}^{\mu}_{(n)} + \log z \, \widetilde{\theta}^{\mu}_{(n)} \right) , \qquad \widetilde{\mathscr{L}} = \sum_n z^n \left( \widetilde{\mathscr{L}}^{(n)} + \log z \, \widetilde{\ell}^{(n)} \right) ,
\end{equation}
equation \eqref{Spin-1 Poincar'e Presymp Radial Eq} becomes
\begin{equation} \label{Spin-1 Poincar'e Presymp Radial Eq order}
    \left(n-D+4\right) \widetilde{\Theta}^{z}_{(n)} + \widetilde{\theta}^{z}_{(n)} \approx \delta \widetilde{\mathscr{L}}^{(n-1)} - \partial_a \widetilde{\Theta}^{a}_{(n-1)} \, .
\end{equation}
Therefore, one can see that the orders $n < D-4$ of $\widetilde{\Theta}^{z}_{(n)}$, which are the ones that come with divergent prefactors in $\Theta^z$, are fixed on-shell to be total derivatives plus total variations that can be removed order by order, while $\widetilde{\Theta}^{z}_{(D-4)}$, which gives the finite order of $\Theta^z$, is undetermined by this equation.  In order to see how the procedure concretely works one has to distinguish even- and odd-dimensional cases. 

\subsubsection{Even dimensions}

In the generic even dimensional diverging case, i.e. for $D \geq 6$, the asymptotic radial expansion of the presymplectic potential radial component is
\begin{equation}
    \Theta^z = \sum_{n=0}^{\frac{D-6}{2}} z^{2n-D+5} \, \widetilde{\Theta}^{z}_{(2n+1)} + \widetilde{\Theta}^{z}_{(D-4)} + \mathcal{O}(z) \, ,
\end{equation}
where the divergent orders are given in terms of the asymptotic solution space of the massless spin-1 field and its field strength by
\begin{equation} \label{spin1 FG generic non log term presymp pot}
    \widetilde{\Theta}^{z}_{(2n+1)} = - \frac{1}{\ell} \sum_{q=0}^{n} F_{za}^{(2(n-q)+1)} \delta A^{a}_{(2q)} \, ,
\end{equation}
while the finite order is
\begin{equation}
    \widetilde{\Theta}^{z}_{(D-4)} = - \frac{1}{\ell} F_{za}^{(D-4)} \delta A^{a}_{(0)} \, .
\end{equation}
According to \eqref{Spin-1 Poincar'e Presymp Radial Eq order}, the divergent orders can be expressed as ambiguities built on $\partial_a \widetilde{\Theta}^{a}_{(2n)}$ and $\delta \widetilde{\mathscr{L}}^{(2n)}$, where
\begin{subequations} \label{spin1 FG generic presymp renorm ambiguities corner}
\begin{align}
    \text{odd }n : \quad
    \begin{split}
    \widetilde{\Theta}^{a}_{(2n)} &= - \ell \sum_{q=0}^{2n-1} \left( {F^a}_{b}^{(2n-2q)} \delta A^{b}_{(2q)} \right) - \frac{1}{\ell} {F^a}_{z}^{(n)} \delta A_z^{(n)} \, ,
    \end{split}\\
    \text{even }n : \quad
    \begin{split}
    \widetilde{\Theta}^{a}_{(2n)} &= - \ell \sum_{q=0}^{2n-1} \left( {F^a}_{b}^{(2n-2q)} \delta A^{b}_{(2q)} \right) - \frac{1}{\ell} \left( {F^a}_{z}^{(n-1)} \delta A_z^{(n+1)} + {F^a}_{z}^{(n+1)} \delta A_z^{(n-1)} \right) ,
    \end{split}
\end{align}
\end{subequations}
and
\begin{subequations}\label{spin1 FG generic presymp renorm ambiguities bdy}
\begin{align}
    \text{odd }n : \quad
    &\widetilde{\mathscr{L}}^{(2n)} = - \frac{\ell}{4}  \sum_{q=0}^{2n-1} F^{ab}_{(2n-2q)} F_{ab}^{(2q)} + \frac{1}{2\ell} {F^a}_{z}^{(n)} F_{za}^{(n)} \, ,\\
    \text{even }n : \quad
    &\widetilde{\mathscr{L}}^{(2n)} = - \frac{\ell}{4} \sum_{q=0}^{2n-1} F^{ab}_{(2n-2q)} F_{ab}^{(2q)} + \frac{1}{\ell} {F^a}_{z}^{(n+1)} F_{za}^{(n-1)} \, .
\end{align}
\end{subequations}
The counterterm is therefore
\begin{equation}
    \Theta^z_{\text{ct}} = - \sum_{n=0}^{\frac{D-6}{2}} z^{2n-D+5} \left( \partial_a \widetilde{\Theta}^{a}_{(2n)} - \delta \widetilde{\mathscr{L}}^{(2n)} \right) ,
\end{equation}
and the renormalized presymplectic potential is
\begin{equation}
    \Theta^z_{\text{ren}} = \lim_{z\to 0} \left( \Theta^z + \Theta^z_{\text{ct}} \right) = \widetilde{\Theta}^{z}_{(D-4)} = - \frac{1}{\ell} F_{za}^{(D-4)} \delta A^{a}_{(0)} \, .
\end{equation}
This procedure involves a choice of the gauge fixation of the $\mathcal{A}-$field in the asymptotic solution spaces. However, in the way we have written it above, the systematics is gauge independent, as is the final result for the finite renormalized part of the potential.

\subsubsection{Odd dimensions}

For $D \geq 5$, the divergent terms of the presymplectic potential read
\begin{equation}
    \Theta^z = \sum_{n=0}^{\frac{D-7}{2}} z^{2n-D+5} \, \widetilde{\Theta}^{z}_{(2n+1)} + \log z \, \widetilde{\theta}^{z}_{(D-4)} + \mathcal{O}(1) \, ,
\end{equation}
where, in terms of the asymptotic solution spaces, the non-logarithmic terms are given by \eqref{spin1 FG generic non log term presymp pot} and the logarithmic one by
\begin{equation}
    \widetilde{\theta}^{z}_{(D-4)} = - \frac{1}{\ell} \widetilde{F}_{za}^{(D-4)} \delta A^{a}_{(0)} \, .
\end{equation}
Using the radial equation \eqref{Spin-1 Poincar'e Presymp Radial Eq order} and the expressions \eqref{spin1 FG generic presymp renorm ambiguities corner}-\eqref{spin1 FG generic presymp renorm ambiguities bdy}, we can cancel the above divergences with the following counterterm
\begin{equation} \label{counterterm odd}
    \Theta^z_{\text{ct}} = - \sum_{n=0}^{\frac{D-7}{2}} z^{2n-D+5} \left( \partial_a \widetilde{\Theta}^{a}_{(2n)} - \delta \widetilde{\mathscr{L}}^{(2n)} \right) + \log z \left( \partial_a \widetilde{\Theta}^{a}_{(D-5)} - \delta \widetilde{\mathscr{L}}^{(D-5)} \right)
\end{equation}
such that the renormalized presymplectic potential takes the form
\begin{equation} \label{presymplectic odd}
    \Theta^z_{\text{ren}} = \lim_{z\to 0} \left( \Theta^z + \Theta^z_{\text{ct}} \right) = \widetilde{\Theta}^{z}_{(D-4)} = - \frac{1}{\ell} \sum_{n=0}^{\frac{D-5}{2}} F_{za}^{(D-4-2n)} \delta A^{a}_{(2n)} \, .
\end{equation}
Although this finite order is not subject to the renormalization equation, we can still add finite ambiguities to it. In particular, if we add the following corner term
\begin{equation} \label{spin1 FG generic presymp renorm ambiguities corner gauge dep}
    \Theta^{\text{corner}}_c = - \frac{\ell^{D-4} \sum_{n=1}^{\frac{D-5}{2}} \frac{1}{n}}{2^{D-4} \left[ \left( \frac{D-5}{2} \right) ! \right]^2} \left( \partial^b \Box^{\frac{D-7}{2}} F_{ab}^{(0)} \partial_c \delta A^a_{(0)} - \frac{2}{D-3} \partial^b \Box^{\frac{D-7}{2}} F_{cb}^{(0)} \partial \cdot \delta A^{(0)} \right) ,
\end{equation}
we obtain
\begin{equation}
    \Theta^z_{\text{ren}} = - \frac{1}{\ell} F_{za}^{(D-4)} \delta A^a_{(0)} - \frac{\ell^{D-4}}{2^{D-4} \left[ \left( \frac{D-5}{2} \right) ! \right]^2} \left( \sum_{n=1}^{\frac{D-5}{2}} \frac{1}{n} \right) \partial^b \Box^{\frac{D-5}{2}} F_{ab}^{(0)} \delta A^a_{(0)} \, .
\end{equation}
Note that, despite involving a gauge choice for the spin-one field, the final result and the entire procedure above are gauge-independent. The only concrete effect of the Lorenz gauge that we employed above is in the factor $\left(\frac{2}{D-3}\right)$ of \eqref{spin1 FG generic presymp renorm ambiguities corner gauge dep} that in the radial gauge, for instance, would be equal to $1$. We can further enhance the last expression with the addition of the following terms
\begin{subequations}
    \begin{align}
        &\Theta_{\text{bdy}} = \frac{\ell^{D-4}}{2^{D-2} \left[ \left( \frac{D-5}{2} \right) ! \right]^2}  \left( \sum_{n=1}^{\frac{D-5}{2}} \frac{1}{n} \right) F^{ab}_{(0)} \, \Box^{\frac{D-5}{2}} F_{ab}^{(0)} \, ,\\
        &\Theta^{\text{corner}}_{c} = \frac{\ell^{D-4}}{2^{D-4} \left[ \left( \frac{D-5}{2} \right) ! \right]^2} \left( \sum_{n=1}^{\frac{D-5}{2}} \frac{1}{n} \right) \left( \Box^{\frac{D-5}{2}} F_{ac}^{(0)} \delta A^a_{(0)} + \frac{D-5}{4} \partial_c \Box^{\frac{D-7}{2}} F_{ab}^{(0)} \delta F^{ab}_{(0)} \right) ,
    \end{align}
\end{subequations}
to end up with the expected result
\begin{equation}
    \Theta^z_{\text{ren}} = - \frac{1}{\ell} \delta A^{a}_{(0)} {F}_{za}^{(D-4)} \, .
\end{equation}

Let us consider explicitly the case of $D=7$, so as to also highlight the role played by different gauge choices, for the two instances of the Lorenz gauge and the radial gauge.

\paragraph{Lorenz gauge} In this case, the asymptotic radial expansion of $\Theta^z$ is
\begin{equation} \label{asymptExp Theta z spin 1 D=7 Poincar'e Lorenz Radial gauge}
    \Theta^z = \frac{1}{z^2} \widetilde{\Theta}^{z}_{(1)} + \log z \, \widetilde{\theta}^{z}_{(3)} + \widetilde{\Theta}^{z}_{(3)} + \mathcal{O}(z^2) \, ,
\end{equation}
where the diverging orders are
\begin{equation} \label{asymptExp Theta z spin 1 D=7 Poincar'e Lorenz Radial gauge diverging orders}
    \widetilde{\Theta}^{z}_{(1)} = - \frac{\ell}{2} \partial^b F_{ab}^{(0)} \delta A^{a}_{{(0)}} \, , \qquad \widetilde{\theta}^{z}_{(3)} = \frac{\ell^3}{4} \partial^b \Box F_{ab}^{(0)} \delta A^{a}_{(0)} \, ,
\end{equation}
and where in particular the solution space of $\mathcal{A}_{\mu}$ is such that  
\begin{equation} \label{asymptExp A field spin 1 D=7 Poincar'e Lorenz gauge}
    \mathcal{A}_{a} = A_{a}^{(0)} + z^2 A_{a}^{(2)} + \mathcal{O}(z^4) \, , \qquad A_a^{(2)} = \frac{\ell^2}{8} \left( 2 \Box A_a^{(0)} - \partial_a \partial \cdot A^{(0)} \right) ,
\end{equation}
with $A_{a}^{(0)}$ a free function of $x^a$.
The divergences are cancelled thanks to the following counterterm 
\begin{equation} \label{asymptExp Theta z spin 1 D=7 Poincar'e Lorenz Radial gauge counterterm1}
    \Theta^z_{\text{ct}} = - \frac{1}{z} \left( \partial_a \widetilde{\Theta}^{a}_{(0)} - \delta \widetilde{\mathscr{L}}^{(0)} \right) + \frac{1}{z^3} \log z \left( \partial_a \widetilde{\Theta}^{a}_{(2)} - \delta \widetilde{\mathscr{L}}^{(2)} \right)
\end{equation}
since the asymptotic renormalization equation \eqref{Spin-1 Poincar'e Presymp Radial Eq order} yields
\begin{equation} \label{asymptExp Theta z spin 1 D=7 Poincar'e Lorenz Radial gauge counterterm2}
    \widetilde{\Theta}^{z}_{(1)} \approx \frac{1}{2} \left( \partial_a \widetilde{\Theta}^{a}_{(0)} - \delta \widetilde{\mathscr{L}}^{(0)} \right) \, , \qquad \widetilde{\theta}^{z}_{(3)} \approx \delta \widetilde{\mathscr{L}}^{(2)} - \partial_a \widetilde{\Theta}^{a}_{(2)} \, ,
\end{equation}
where
\begin{equation} \label{asymptExp Theta z spin 1 D=7 Poincar'e Lorenz Radial gauge counterterm3}
    \widetilde{\Theta}^{a}_{(0)} = - \ell  {F^{ab}_{(0)}} \delta A_b^{(0)} \, , \qquad \widetilde{\mathscr{L}}^{(0)} = - \frac{\ell}{4} F^{ab}_{(0)} F_{ab}^{(0)} \, ,
\end{equation}
and
\begin{subequations} \label{asymptExp Theta z spin 1 D=7 Poincar'e Lorenz Radial gauge counterterm4}
    \begin{align}
        \widetilde{\Theta}^{a}_{(2)} &= \frac{\ell^3}{32} \left[ \partial_b F^{ab}_{(0)} \partial \cdot \delta A^{(0)} + 4 F^{ab}_{(0)} \partial_b \partial \cdot \delta A^{(0)} - 8 \left( \Box F^{ab}_{(0)} \delta A_b^{(0)} + F^{ab}_{(0)} \Box \delta A_b^{(0)} \right) \right] ,\\
        \widetilde{\mathscr{L}}^{(2)} &= - \frac{\ell^3}{8} \left( \partial^b F_{ab}^{(0)} \partial_c F^{ac}_{(0)} + F^{ab}_{(0)} \Box F_{ab}^{(0)} \right) .
    \end{align}
\end{subequations}
Therefore, the renormalized presymplectic potential \eqref{presymplectic odd} is
\begin{equation} \label{asymptExp Theta z spin 1 D=7 Poincar'e Lorenz gauge ren}
    \Theta^z_{\text{ren}} = - \frac{1}{\ell} {F}_{za}^{(3)} \delta A^{a}_{(0)} - \frac{\ell^3}{8} \partial^b F_{ab}^{(0)} \Box \delta A^{a}_{(0)} + \frac{\ell^3}{16} \partial^b F_{ab}^{(0)} \partial^a \partial \cdot \delta A^{(0)} \, .
\end{equation}
If we add the following corner and boundary terms
\begin{subequations}
    \begin{align}
    &\Theta_c^{\text{corner}} = \frac{\ell^3}{16} \left( \partial^b F_{cb}^{(0)} \partial \cdot \delta A^{(0)} - 2 \partial^b F_{ab}^{(0)} \partial_c \delta A^a_{(0)} + 2 \delta A^a_{(0)} \Box F_{ac}^{(0)} + \partial_c F_{ab}^{(0)} \delta F^{ab}_{(0)} \right) ,\\
    &\Theta_{\text{bdy}} = \frac{\ell^3}{32} F^{ab}_{(0)} \Box F_{ab}^{(0)} \, ,
    \end{align}
\end{subequations}
the renormalized presymplectic potential reads
\begin{equation} \label{asymptExp Theta z spin 1 D=7 Poincar'e Lorenz gauge ren2}
    \Theta^z_{\text{ren}} = - \frac{1}{\ell} \delta A^{a}_{(0)} {F}_{za}^{(3)} \, .
\end{equation}
and matches the result 
we obtained by renormalizing the variational principle. Note that the ambiguity in $\delta$-exact must be justified by a boundary term to add to the bulk action.

\paragraph{Radial gauge}

In the radial gauge $\mathcal{A}_{z}=0$, the solution space of the $\mathcal{A}-$field takes the form
\begin{equation} \label{asymptExp A field spin 1 D=7 Poincar'e Radial gauge}
    \mathcal{A}_{a} = A_{a}^{(0)} + z^2 A_{a}^{(2)} + \mathcal{O}(z^4) \, , \qquad A_a^{(2)} = \frac{\ell^2}{4} \left( \Box A_a^{(0)} - \partial_a \partial \cdot A^{(0)} \right) = \frac{\ell^2}{4} \partial^b F_{ba}^{(0)} \, ,
\end{equation}
where $A_{a}^{(0)}$ is an arbitrary function of the boundary coordinates. The renormalization is identical to what we performed for the Lorenz gauge. The radial component of the presymplectic potential then radially expands as in \eqref{asymptExp Theta z spin 1 D=7 Poincar'e Lorenz Radial gauge}, where the on-shell values of the diverging orders are given in \eqref{asymptExp Theta z spin 1 D=7 Poincar'e Lorenz Radial gauge diverging orders}. Using \eqref{Spin-1 Poincar'e Presymp Radial Eq order}, we have shown that the latter can be cancelled if we add the counterterm \eqref{asymptExp Theta z spin 1 D=7 Poincar'e Lorenz Radial gauge counterterm1}--\eqref{asymptExp Theta z spin 1 D=7 Poincar'e Lorenz Radial gauge counterterm4} to the presymplectic potential, except that in the present case
\begin{equation}
    \widetilde{\Theta}^{a}_{(2)} = \frac{\ell^3}{4} \left( 2 F^{ab}_{(0)} \partial_b \partial \cdot \delta A^{(0)} - \Box F^{ab}_{(0)} \delta A_b^{(0)} - F^{ab}_{(0)} \Box \delta A_b^{(0)} \right) .
\end{equation}
This leads us to the on-shell renormalized symplectic structure
\begin{equation}
    \Theta^z_{\text{ren}} \approx - \frac{1}{\ell} {F}_{za}^{(3)} \delta A^{a}_{(0)} - \frac{\ell^3}{8} \partial^b F_{ab}^{(0)} \Box \delta A^{a}_{(0)} + \frac{\ell^3}{8} \partial^b F_{ab}^{(0)} \partial^a \partial \cdot \delta A^{(0)}
\end{equation}
which differs from \eqref{asymptExp Theta z spin 1 D=7 Poincar'e Lorenz gauge ren} obtained in the Lorenz gauge by a pure corner term
\begin{equation}
    \partial^a \left( - \frac{\ell^3}{16} \partial^b F_{ab}^{(0)} \partial \cdot \delta A^{(0)} \right) .
\end{equation}
We can then conclude that we obtain the same result in both gauges. Note that we can still fix the ambiguities of the presymplectic potential for the finite order. Hence, if we add the same finite boundary and corner terms to the renormalized potential as in the Lorenz gauge, it can finally be rewritten as \eqref{asymptExp Theta z spin 1 D=7 Poincar'e Lorenz gauge ren2}.

\section{Spin-one fields in Bondi coordinates} \label{sec:spin1_bondi}

We want to determine the renormalized asymptotic spin-one charges in Bondi coordinates~\ref{app:Bondi}. One motivation is that in this framework computing the flat limit of the charge becomes essentially trivial. However, in Bondi coordinates the analysis in arbitrary dimensions turns out to be more involved than in the Poincar\'e patch and we will detail only a few specific cases.

\subsection{Solution space}

The equations of motion for the field strength read
\begin{subequations} \label{Bondi-Spin1-FieldF-eom}
    \begin{align}
        0 &= \left( \partial_r + \frac{D-2}{r} \right) \mathcal{F}_{ur} - \frac{1}{r^2} \partial^i \mathcal{F}_{ir} \, ,\\
        0 &= \partial_u \mathcal{F}_{ru} - \frac{1}{r^2} \partial^i \mathcal{F}_{iu} + \left( \frac{1}{r^2} + \frac{1}{\ell^2} \right) \partial^i \mathcal{F}_{ir} \, ,\\
        0 &= \frac{1}{r^2} \left( \partial_r + \frac{D-4}{r} \right) \left( \mathcal{F}_{ri} - \mathcal{F}_{ui} \right) + \frac{1}{\ell^2} \left( \partial_r + \frac{D-2}{r} \right) \mathcal{F}_{ri} - \frac{1}{r^2} \partial_u \mathcal{F}_{ri} - \frac{1}{r^4} \partial^j \mathcal{F}_{ij} \, .
    \end{align}
\end{subequations}
Assuming an asymptotic radial expansion of the form
\begin{equation} \label{Bondi F spin-1 generic asymp exp}
    \mathcal{F}_{\mu\nu}(u,r,x^i) = \sum_{n} r^{-n} \left( F_{\mu\nu}^{(n)}(u,x^i) + \log r \, \widetilde{F}_{\mu\nu}^{(n)}(u,x^i) \right) ,
\end{equation}
the above equations of motion yield the following recursive relations
\begin{subequations} \label{Bondi-Spin1-FieldF-eom-symp}
    \begin{align}
        0 &= \left(D-n-2\right) F_{ur}^{(n)} + \widetilde{F}_{ur}^{(n)} - \partial^i F_{ir}^{(n-1)} \, ,\label{Bondi-Spin1-FieldF-eom1-symp}\\
        0 &= \partial_u F_{ru}^{(n)} + \partial^i \left( F_{ir}^{(n-2)} - F_{iu}^{(n-2)} \right) + \frac{1}{\ell^2} \partial^i F_{ir}^{(n)} \, ,\label{Bondi-Spin1-FieldF-eom2-symp}\\
        \begin{split} \label{Bondi-Spin1-FieldF-eom3-symp}
        0 &= (D-n-4) \left(F_{ri}^{(n)} - F_{ui}^{(n)}\right) + \left(\widetilde{F}_{ri}^{(n)} - \widetilde{F}_{ui}^{(n)}\right) + \frac{1}{\ell^2} \left(D-n-4\right) F_{ri}^{(n+2)} \\
        &\quad + \frac{1}{\ell^2} \widetilde{F}_{ri}^{(n+2)} - \partial_u \, F_{ri}^{(n+1)} - \partial^j F_{ij}^{(n-1)} \, ,
        \end{split}
    \end{align}
\end{subequations}
while the Bianchi identities provide the additional constraints
\begin{subequations} \label{Bondi-Spin1-FieldF-Bianchi-symp}
    \begin{align}
        &\partial_u F_{ir}^{(n)} - \partial_i F_{ur}^{(n)} = - (n-1) F_{iu}^{(n-1)} + \widetilde{F}_{iu}^{(n-1)} \, ,\label{Bondi-Spin1-FieldF-Bianchi1-symp}\\
        &\partial_i F_{uj}^{(n)} - \partial_j F_{ui}^{(n)} = \partial_u F_{ij}^{(n)} \, ,\label{Bondi-Spin1-FieldF-Bianchi2-symp}\\
        &\partial_i F_{rj}^{(n)} - \partial_j F_{ri}^{(n)} = - (n-1) F_{ij}^{(n-1)} + \widetilde{F}_{ij}^{(n-1)} \, ,\label{Bondi-Spin1-FieldF-Bianchi3-symp}\\
        &\partial_i F_{kj}^{(n)} - \partial_j F_{ki}^{(n)} = \partial_k F_{ij}^{(n)} \, .\label{Bondi-Spin1-FieldF-Bianchi4-symp}
    \end{align}
\end{subequations}
The relations governing the logarithmic terms are obtained similarly.  One can streamline a bit the recursive relation for the radial orders $F_{ij}^{(n)}$ by injecting \eqref{Bondi-Spin1-FieldF-Bianchi2-symp}, \eqref{Bondi-Spin1-FieldF-Bianchi3-symp} and \eqref{Bondi-Spin1-FieldF-Bianchi4-symp} into the antisymmetric spatial derivative of \eqref{Bondi-Spin1-FieldF-eom3-symp}
\begin{equation}
    \begin{split}
    0 &= \frac{1}{\ell^2} (D-n-4) (n+1) F_{ij}^{(n+1)} - \frac{1}{\ell^2} (D-2n-5) \widetilde{F}_{ij}^{(n+1)} + (D-2n-4) \partial_u F_{ij}^{(n)}\\
    &\quad  + 2 \partial_u \widetilde{F}_{ij}^{(n)} - \left( \Delta - (D-n-4) (n-1) \right) F_{ij}^{(n-1)} - (D-2n-3) \widetilde{F}_{ij}^{(n-1)} \, ,
    \end{split}
    \label{Spin1-Bondi-FieldF-Eq1ToTakeCare-SolutionSpace-symp}
\end{equation}
where we recall that $\Delta$ is the Laplacian operator w.r.t. $\gamma_{ij}$. The combination of \eqref{Bondi-Spin1-FieldF-eom1-symp} and \eqref{Bondi-Spin1-FieldF-Bianchi1-symp} leads to, for $n \neq \{0 , D-3\}$,
\begin{equation}
    F_{iu}^{(n)} = - \frac{1}{n} \left( \partial_u F_{ir}^{(n+1)} - \frac{\partial_i \partial^j F_{jr}^{(n)}}{D-n-3} \right) + \frac{1}{n} \widetilde{F}_{iu}^{(n)} \, ,
    \label{Spin1-Bondi-FieldF-Eq4ToTakeCare-SolutionSpace-symp}
\end{equation}
and yields a recursive relation for $F_{ir}^{(n)}$ in terms of $F_{ij}^{(n)}$ when injected into \eqref{Bondi-Spin1-FieldF-eom3-symp}
\begin{equation}\label{Spin1-Bondi-FieldF-Eq2ToTakeCare-SolutionSpace-symp}
    \begin{split}
    0 &= \frac{1}{\ell^2} (D-n-4) F_{ir}^{(n+2)} + \frac{1}{\ell^2} \widetilde{F}_{ir}^{(n+2)} - \partial_u F_{ir}^{(n+1)} + (D-n-4) F_{ir}^{(n)} \\
    &\quad + \left( \widetilde{F}_{ir}^{(n)} - \widetilde{F}_{iu}^{(n)} \right) + \frac{D-n-4}{n} \left( \partial_u F_{ir}^{(n+1)} - \frac{\partial_i \partial^j F_{jr}^{(n)}}{D-n-3} - \widetilde{F}_{iu}^{(n)} \right) + \partial^j F_{ij}^{(n-1)} \, .
    \end{split}
\end{equation}

In terms of $\mathcal{A}_{\mu}$, in the radial gauge $\mathcal{A}_r = 0$, the equations of motion \eqref{Bondi-Spin1-FieldF-eom} become
\begin{subequations} \label{Radial-equationofmotion-Bondi-spin1}
    \begin{align}
    0 &= \left[ r^2 (\partial_r - \partial_u) \partial_r + r (D-2) \partial_r + \Delta \right] \mathcal{A}_u - \partial_u D \cdot A + \frac{r^3}{\ell^2} \left( r \partial_r + D - 2 \right) \partial_r \mathcal{A}_u \, ,
    \label{Radial-equationofmotion-Bondi-spin1-u-symp}\\
    0 &= r \left( r \partial_r + D - 2 \right) \partial_r \mathcal{A}_u - \partial_r D \cdot A \, ,\label{Radial-equationofmotion-Bondi-spin1-r-symp}\\
    \begin{split} \label{Radial-equationofmotion-Bondi-spin1-i-symp}
    0 &= r \left( r \partial_r + D - 4 \right) ( \partial_i \mathcal{A}_u - \partial_u \mathcal{A}_i ) -  D_i D \cdot A - r^2 \partial_u \partial_r \mathcal{A}_i + \frac{r^3}{\ell^2} \left( r \partial_r + D - 2 \right) \partial_r \mathcal{A}_i\\
    & \quad + \left[ r^2 \partial_r^2 + (D-4) r \partial_r + \Delta \right] \mathcal{A}_i \, .
    \end{split}
    \end{align}
\end{subequations}
In analogy with \eqref{Bondi F spin-1 generic asymp exp}, assuming
\begin{equation}
\mathcal{A}_\mu(u,r,x^i) = \sum_n r^{-n} \left( A_\mu^{(n)}(u,x^i) + \log r \, \widetilde{A}_\mu^{(n)}(u,x^i) \right) \, ,
\end{equation}
we obtain
\begin{subequations} \label{Radial-equationofmotion-Bondi-spin1-symp-AsymptExp-r}
    \begin{align}
        \begin{split}
        0 &= \frac{1}{\ell^2}(n+1) (n-D+4) A_u^{(n+1)} + \left[ \Delta + (n-1) \, (n-D+2) \right] A_u^{(n-1)} + n \partial_u A_u^{(n)}\\
        &\quad - \partial_u D \cdot A^{(n-1)} - \partial_u \widetilde{A}_u^{(n)} + (D - 2n - 1) \widetilde{A}_u^{(n-1)} + \frac{1}{\ell^2} (D - 2n - 5) \widetilde{A}_u^{(n+1)} \, ,
        \end{split}
        \label{Radial-equationofmotion-Bondi-spin1-u-symp-AsymptExp-r}\\
        \begin{split}
        0 &= (n-1) (n-D+2) A_u^{(n-1)} + (n-2) D \cdot A^{(n-2)} + (D - 2n - 1) \widetilde{A}_u^{(n-1)}\\
        &\quad - D \cdot \widetilde{A}^{(n-2)} \, ,
        \end{split}
        \label{Radial-equationofmotion-Bondi-spin1-r-symp-AsymptExp-r}\\
        \begin{split}
        0 &= (2n - D + 4) \partial_u A_i^{(n)} + (D-n-4) \partial_i A_u^{(n)} - D_i D \cdot A^{(n-1)} - 2 \partial_u \widetilde{A}_i^{(n)} + \partial_i \widetilde{A}_u^{(n)}\\
        &\quad + (D-2n-1) \widetilde{A}_i^{(n-1)} + \left[ \Delta + (n-1) (n-D+4) \right] A_i^{(n-1)}\\
        &\quad + \frac{1}{\ell^2} (n+1) (n-D+4) A_i^{(n+1)} + \frac{1}{\ell^2} ( D - 2n - 5 ) \widetilde{A}_i^{(n+1)} \, .
        \label{Radial-equationofmotion-Bondi-spin1-i-symp-AsymptExp-r}
        \end{split}
    \end{align}
\end{subequations}

\subsection{Holographic renormalization}

We introduce the regulated on-shell action
\begin{equation}
S_{\text{reg}}^{(\epsilon)} \approx - \frac{1}{2} \int_{r \leq \epsilon} \text{d}^{D}x \, \partial_\mu \left( \sqrt{-g} \mathcal{A}_\nu \mathcal{F}^{\mu\nu} \right) .
\end{equation}
Since the orientation of the boundary is given by $n_\mu = \delta_\mu^r$, the above action reduces to
\begin{equation}
S_{\text{reg}}^{(\epsilon)} = - \frac{1}{2} \int_{r = \epsilon} \text{d}^{D-1}x \, \epsilon^{D-2} \sqrt{-\gamma} \left[ \mathcal{A}_u \mathcal{F}_{ur} - \frac{1}{\epsilon^2} \mathcal{A}^i \mathcal{F}_{ui} + \left( \frac{1}{\epsilon^2} + \frac{1}{\ell^2} \right) \mathcal{A}^i \mathcal{F}_{ri} \right] .
\end{equation}
Let us show how to obtain its renormalized on-shell variation in two explicit instances.

\subsubsection{$D = 6$} 
By solving the equations \eqref{Bondi-Spin1-FieldF-Bianchi-symp}--\eqref{Spin1-Bondi-FieldF-Eq2ToTakeCare-SolutionSpace-symp} with the additional that all positive powers of $r$ vanish, one obtains the asymptotic solution space for $D=6$
\begin{subequations} \label{Spin 1 Bondi ASP F D=6}
    \begin{align}
        &\mathcal{F}_{ij} = F_{ij}^{(0)} + \tfrac{1}{r} F_{ij}^{(1)} + \tfrac{1}{r^2} F_{ij}^{(2)} + \tfrac{1}{r^3} F_{ij}^{(3)} + \mathcal{O}(\tfrac{1}{r^4}) \, ,\\
        &\mathcal{F}_{ir} = \tfrac{1}{r^2} F_{ir}^{(2)} + \tfrac{1}{r^3} F_{ir}^{(3)} + \tfrac{1}{r^4} F_{ir}^{(4)} + \mathcal{O}(\tfrac{1}{r^5}) \, ,\\
        &\mathcal{F}_{ur} = \tfrac{1}{r^3} F_{ur}^{(3)} + \tfrac{1}{r^4} F_{ur}^{(4)} + \mathcal{O}(\tfrac{1}{r^5}) \, ,\\
        &\mathcal{F}_{iu} = F_{iu}^{(0)} + \tfrac{1}{r} F_{iu}^{(1)} + \mathcal{O}(\tfrac{1}{r^2}) \, ,
    \end{align}
\end{subequations}
where $F_{ij}^{(0)}$, $F_{ij}^{(3)}$ and $F_{iu}^{(0)}$ are arbitrary functions of $(u,x^i)$, while for instance
\begin{equation}
    \begin{split}
        &F_{ij}^{(1)} = - \ell^2 \partial_u F_{ij}^{(0)} \, , \qquad F_{ij}^{(2)} = \tfrac{\ell^2}{2} \Delta F_{ij}^{(0)} \, , \qquad F_{ir}^{(2)} = \ell^2 F_{iu}^{(0)} \, ,\\
        &F_{ir}^{(3)} = - \ell^2 \partial^j F_{ij}^{(0)} \, , \qquad F_{ur}^{(3)} = \ell^2 \partial^i F_{iu}^{(0)} \, , \qquad F_{iu}^{(1)} = - \ell^2 \partial_u F_{iu}^{(0)} \, ,
    \end{split}
\end{equation}
and
\begin{equation} \label{Additional constraint for conversation - Field Strength - D=6}
    0 = \frac{1}{\ell^2} \partial^i F_{ir}^{(4)} + \ell^2 \partial^i F_{iu}^{(0)} - \frac{\ell^2}{2} \partial^i \partial^j \partial_i F_{ju}^{(0)} - \partial_u F_{ur}^{(4)} \, .
\end{equation}
In the regulated on-shell action we can highlight the divergent part
\begin{equation}
        S_{\text{reg}}^{(\epsilon)} = - \frac{1}{2} \int_{r=\epsilon} \text{d}^5x \, \epsilon \sqrt{-\gamma} \left[ \mathcal{A}^i \partial^j \mathcal{F}_{ij} + \ell^2 \left( \mathcal{A}_u \partial^i - \mathcal{A}^i \partial_u \right) \mathcal{F}_{iu} \right] + \mathcal{O}(1) \, ,
\end{equation}
where we used
\begin{equation}
    F_{ij}^{(0)} = \left(1 + \frac{\ell^2}{r} \partial_u \right) \mathcal{F}_{ij} + \mathcal{O}\left(r^{-2}\right) , \qquad F_{iu}^{(0)} = \left(1 + \frac{\ell^2}{r} \partial_u \right) \mathcal{F}_{iu} + \mathcal{O}\left(r^{-2}\right) .
\end{equation}
We then add the following counterterm
\begin{equation}
        S_{\text{ct}}^{(\epsilon)} = \frac{1}{4} \int_{r=\epsilon} \text{d}^5x \, \epsilon \, \sqrt{-\gamma} \left( \mathcal{F}^{ij} \mathcal{F}_{ij} - 2 \ell^2 \mathcal{F}^{i}{}_u \mathcal{F}_{iu} \right) ,
\end{equation}
where we cancelled a boundary term by means of a corner term
\begin{equation}
    S_{\text{corner}}^{(\epsilon)} = \frac{1}{2} \int_{r=\epsilon} \text{d}^5x \, \epsilon \, \sqrt{-\gamma} \left[ \partial^i \left( \mathcal{A}^j \mathcal{F}_{ji} + \ell^2 \mathcal{A}_u \mathcal{F}_{iu} \right) - \ell^2 \partial_u \left( \mathcal{A}^i \mathcal{F}_{iu} \right) \right] .
\end{equation}
Therefore, the subtracted action $S_{\text{sub}}^{(\epsilon)} = S_{\text{reg}}^{(\epsilon)} + S_{\text{ct}}^{(\epsilon)}$ takes the following form
\begin{equation}
    S_{\text{sub}}^{(\epsilon)} = \int_{r \leq \epsilon} \text{d}^{6}x \left( - \frac{\sqrt{-g}}{4} \mathcal{F}_{\mu\nu} \mathcal{F}^{\mu\nu} \right) + \frac{1}{4} \int_{r=\epsilon} \text{d}^5x \, \epsilon \, \sqrt{-\gamma} \left( \mathcal{F}^{ij} \mathcal{F}_{ij} - 2 \ell^2 \mathcal{F}^{i}{}_u \mathcal{F}_{iu} \right) .
\end{equation}
By varying the latter on-shell and taking the limit $\epsilon \to 0$, one obtains
\begin{equation} \label{delta Sren D=6 any gauge}
    \delta S_{\text{ren}} \approx - \int \text{d}^5x \, \sqrt{-\gamma} \left[ \delta \mathcal{A}_u F_{ur}^{(4)} - \frac{1}{\ell^2} \delta \mathcal{A}^i F_{ir}^{(4)} + \frac{\ell^2}{2} \delta \mathcal{A}^i \left( \partial^j \partial_u F_{ij}^{(0)} - 2 F_{iu}^{(0)} + \partial_i \partial^j F_{ju}^{(0)} \right) \right] ,
\end{equation}
which becomes, when evaluated along a gauge parameter $\lambda$ and exploiting \eqref{Additional constraint for conversation - Field Strength - D=6},
\begin{equation} \label{charge6B}
    \begin{split}
        \delta_\lambda S_{\text{ren}} &= - \int \text{d}^5x \, \sqrt{-\gamma} \bigg\{ \partial_u \left( \lambda F_{ur}^{(4)} \right) - \tfrac{1}{2 \ell^2} \partial^i \bigg[ \lambda \bigg( 2 F_{ir}^{(4)} - \ell^4 \Big( \partial^j \partial_u F_{ij}^{(0)} - 2 F_{iu}^{(0)}\\
        &\quad + \partial_i \partial^j F_{ju}^{(0)} \Big) \bigg) \bigg] \bigg\} \, .
    \end{split}
\end{equation}
Note that the square bracket in the previous equation drops out, since the integral of its divergence on the sphere vanishes. As a result, we obtain the charge
\begin{equation}\label{QBondi6}
Q^{(\lambda)} = -\int \text{d}^4x \, \sqrt{-\gamma} \, \lambda \, F_{ur}^{(4)}\,.
\end{equation}

\subsubsection{$D=5$} Similarly to the previous case, one determines the asymptotic solution space
\begin{subequations}
    \begin{align}
        &\mathcal{F}_{ij} = F_{ij}^{(0)} + \tfrac{1}{r} F_{ij}^{(1)} + \tfrac{1}{r^2} \left( F_{ij}^{(2)} + \log r \, \widetilde{F}_{ij}^{(2)} \right) + \mathcal{O}(\tfrac{1}{r^3}) \, ,\\
        &\mathcal{F}_{ir} = \tfrac{1}{r^2} F_{ir}^{(2)} + \tfrac{1}{r^3} \left( F_{ir}^{(3)} + \log r \, \widetilde{F}_{ir}^{(3)} \right) + \mathcal{O}(\tfrac{1}{r^4}) \, ,\\
        &\mathcal{F}_{ur} = \tfrac{1}{r^3} \left( F_{ur}^{(3)} + \log r \, \widetilde{F}_{ur}^{(3)} \right) + \mathcal{O}(\tfrac{1}{r^4}) \, ,\\
        &\mathcal{F}_{iu} = F_{iu}^{(0)} + \mathcal{O}(\tfrac{1}{r}) \, ,
    \end{align}
\end{subequations}
where the free data are $F_{ij}^{(0)}$, $F_{ij}^{(2)}$ and $F_{iu}^{(0)}$, and
\begin{equation}
    0 = \frac{1}{\ell^2} \partial^i F_{ir}^{(3)} + \ell^2 \partial^i \partial_u F_{iu}^{(0)} - \partial_u F_{ur}^{(3)} \, .
    \label{Additional constraint for conversation - Field Strength - D=5}
\end{equation}
We shall omit further relations among the coefficients that are not useful to our purposes. Using the inverse expansion
\begin{equation}
    F_{ij}^{(0)} = \mathcal{F}_{ij} + \mathcal{O}\left(r^{-1}\right) , \qquad F_{iu}^{(0)} = \mathcal{F}_{iu} + \mathcal{O}\left(r^{-1}\right) ,
\end{equation}
and adding the corner action
\begin{equation}
    S_{\text{corner}}^{(\epsilon)} = \frac{1}{2} \int_{r=\epsilon} \text{d}^4x \, \log \epsilon \, \sqrt{-\gamma} \left[ \partial^i \left( \mathcal{A}^j \mathcal{F}_{ji} + \ell^2 \mathcal{A}_u \mathcal{F}_{iu} \right) - \ell^2 \partial_u \left( \mathcal{A}^i \mathcal{F}_{iu} \right) \right] ,
\end{equation}
one concludes that the counterterm action is
\begin{equation}
    S_{\text{ct}}^{(\epsilon)} = \frac{1}{4} \int_{r=\epsilon} \text{d}^4x \, \log \epsilon \, \sqrt{-\gamma} \left( \mathcal{F}^{ij} \mathcal{F}_{ij} - 2 \ell^2 \mathcal{F}^{i}{}_u \mathcal{F}_{iu} \right) .
\end{equation}
This leads us to the following variation of the renormalized on-shell action
\begin{equation}
    \delta S_{\text{ren}} \approx - \int \text{d}^4x \, \sqrt{-\gamma} \left( \delta \mathcal{A}_u F_{ur}^{(3)} - \frac{1}{\ell^2} \delta \mathcal{A}^i F_{ir}^{(3)} - \ell^2 \delta \mathcal{A}^i \partial_u F_{iu}^{(0)} \right) .
    \label{delta Sren D=5 any gauge}
\end{equation}
If one evaluates the latter along a gauge parameter $\lambda$ and uses \eqref{Additional constraint for conversation - Field Strength - D=5}, it yields
\begin{equation}\label{charge5B}
    \delta_\lambda S_{\text{ren}} = - \int \text{d}^4x \, \sqrt{-\gamma} \left\{ \partial_u \left( \lambda F_{ur}^{(3)} \right) - \frac{1}{\ell^2} \partial^i \left[ \lambda \left( F_{ir}^{(3)} - \ell^4 \partial_u F_{iu}^{(0)} \right) \right] \right\} .
\end{equation}
Again the square bracket drops out, and we obtain the charge
\begin{equation}\label{QBondi5}
Q^{(\lambda)} = -\int \text{d}^3x \, \sqrt{-\gamma} \, \lambda \, F_{ur}^{(3)} \, .
\end{equation}
\subsection{Symplectic renormalization}

As in \eqref{radial thetatilde scalar} and \eqref{radial Ltilde scalar} we factor out the radial off-shell dependence of the latter and of the Lagrangian,
\begin{equation}
    \Theta^\mu = r^{D-2} \sqrt{-\gamma} \, \widetilde{\Theta}^\mu \, , \qquad \mathscr{L}= r^{D-2} \sqrt{-\gamma} \, \widetilde{\mathscr{L}} \, ,
\end{equation}
where
\begin{subequations} \label{Bondi spin 1 presymp pot off shell}
    \begin{align}
        \widetilde{\Theta}^r &= \mathcal{F}_{ru} \delta \mathcal{A}_u + \tfrac{1}{\ell^2} \mathcal{F}_{ir} \delta \mathcal{A}^i + \tfrac{1}{r^2} (\mathcal{F}_{ui} - \mathcal{F}_{ri}) \delta \mathcal{A}^i , \label{Bondi spin 1 presymp pot off shell compo r}\\
        \widetilde{\Theta}^u &=  \mathcal{F}_{ur} \delta \mathcal{A}_r + \tfrac{1}{r^2} \mathcal{F}_{ir} \delta \mathcal{A}^i , \label{Bondi spin 1 presymp pot off shell compo u}\\
        \widetilde{\Theta}^i &=  \tfrac{1}{\ell^2} {\mathcal{F}_{r}}^i \delta \mathcal{A}_r - \tfrac{1}{r^2} {\mathcal{F}_{r}}^i \delta \mathcal{A}_u - \tfrac{1}{r^2} ({\mathcal{F}_{u}}^i - {\mathcal{F}_{r}}^i) \delta \mathcal{A}_r - \tfrac{1}{r^4} {\mathcal{F}^{i}}_j \delta A^j , \label{Bondi spin 1 presymp pot off shell compo i}
    \end{align}
\end{subequations}
and
\begin{equation} \label{radial Ltilde scalar}
    \widetilde{\mathscr{L}} = \tfrac{1}{2} \left[ \mathcal{F}_{ur} \mathcal{F}_{ur} - \tfrac{1}{\ell^2} {\mathcal{F}_{r}}^i \mathcal{F}_{ri} + \tfrac{1}{r^2}  {\mathcal{F}_{r}}^i \left( \mathcal{F}_{ui} - \mathcal{F}_{ri} \right) - \tfrac{1}{r^2} {\mathcal{F}_{u}}^i \mathcal{F}_{ri} - \tfrac{1}{r^4} \mathcal{F}^{ij} \mathcal{F}_{ij} \right] ,
\end{equation}
so as to get the asymptotic renormalization equation in the form
\begin{equation} \label{Spin-1 Bondi Presymp Radial Eq}
    \frac{1}{r} \left( r \, \partial_r + D - 2 \right) \widetilde{\Theta}^r \approx \delta \widetilde{\mathscr{L}} - \partial_u \widetilde{\Theta}^u - \partial_i \widetilde{\Theta}^i \, .
\end{equation}
Under the assumption
\begin{equation}
    \widetilde{\Theta}^\mu = \sum_n r^{-n} \left( \widetilde{\Theta}^{\mu}_{(n)} + \log r \, \widetilde{\theta}^{\mu}_{(n)} \right) , \qquad \widetilde{\mathscr{L}} = \sum_n r^{-n} \left( \widetilde{\mathscr{L}}^{(n)} + \log r \, \widetilde{\ell}^{(n)} \right) ,
\end{equation}
equation \eqref{Spin-1 Bondi Presymp Radial Eq} delivers the recursive renormalization relation
\begin{equation} \label{Spin-1 Bondi Presymp Radial Eq order}
    \left(D-2-n\right) \widetilde{\Theta}^{r}_{(n)} + \widetilde{\theta}^{r}_{(n)} \approx \delta \widetilde{\mathscr{L}}^{(n+1)} - \partial_u \widetilde{\Theta}^{u}_{(n+1)} - \partial_i \widetilde{\Theta}^{i}_{(n+1)} \, .
\end{equation}
The latter fixes the divergent orders of the presymplectic potential to be ambiguities, corresponding to $n<D-2$, in the very same spirit of section~\ref{sec:symplectic P}. Let us check the consistency of the procedure for the examples of $D=6$ and $D=5$.

\subsubsection{$D=6$} 
Let us recall that, in addition to \eqref{Spin 1 Bondi ASP F D=6}, we must also consider the solution space of the $\mathcal{A}$-field. This one is obtained by solving the equations \eqref{Radial-equationofmotion-Bondi-spin1-symp-AsymptExp-r} in the radial gauge, which gives for $D=6$
\begin{equation}
        \mathcal{A}_i = A_i^{(0)} + \tfrac{1}{r} A_i^{(1)} + \tfrac{1}{r^2} A_i^{(2)} + \mathcal{O}(\tfrac{1}{r^3}) \, , \qquad \mathcal{A}_u = A_u^{(0)} + \mathcal{O}(\tfrac{1}{r^2}) \, ,
\end{equation}
such that $A_i^{(0)}$ and $A_u^{(0)}$ define arbitrary functions of $(u,x^i)$, whereas
\begin{equation}
    A_{i}^{(1)} = \ell^2 \left( \partial_i A_u^{(0)} - \partial_u A_i^{(0)} \right) , \qquad A_{i}^{(2)} = \frac{\ell^2}{2} \left( \Delta A_i^{(0)} - \partial_i D \cdot A^{(0)} \right) .
\end{equation}
Using \eqref{Bondi spin 1 presymp pot off shell compo r} one gets the asymptotic expansion of the radial presymplectic potential
\begin{equation}
    \Theta^r = r \sqrt{-\gamma} \, \widetilde{\Theta}^r_{(3)} + \sqrt{-\gamma} \, \widetilde{\Theta}^r_{(4)} + \mathcal{O}(\tfrac{1}{r}) \, ,
\end{equation}
where divergent and finite orders are given by
\begin{subequations}
    \begin{align}
        \widetilde{\Theta}^r_{(3)} &= \partial^i F_{ij}^{(0)} \delta A^j_{(0)} + \ell^2 F_{ui}^{(0)} \delta F^i_{(0)u} \, ,\\
        \begin{split}
        \widetilde{\Theta}^r_{(4)} &= \frac{1}{\ell^2} F_{ir}^{(4)} \delta A^i_{(0)} + F_{ru}^{(4)} \delta A_u^{(0)} + \frac{\ell^2}{2} \left[ \partial^j \left( \partial_u F_{ij}^{(0)} + \partial_i F_{uj}^{(0)} \right) \delta A^i_{(0)} + 2 F_{iu}^{(0)} \delta A^i_{(0)} \right]\\
        &\quad + \ell^4 \partial_u F_{iu}^{(0)} \delta F^i_{(0)u} \, .
        \end{split}
    \end{align}
\end{subequations}
Thanks to \eqref{Spin-1 Bondi Presymp Radial Eq order}, we can renormalize the above symplectic structure by adding the following counterterm
\begin{equation}
    \Theta^r_{\text{ct}} = r \sqrt{-\gamma} \left( \partial_u \widetilde{\Theta}^u_{(4)} + \partial_i \widetilde{\Theta}^i_{(4)} - \delta \widetilde{\mathscr{L}}_{(4)} \right) ,
\end{equation}
where
\begin{subequations}
    \begin{align}
        \widetilde{\Theta}^u_{(4)} &= \ell^2 F_{iu}^{(0)} \delta A^i_{(0)} \, ,\\
        \widetilde{\Theta}^i_{(4)} &= - F^{i}_{(0)j} \delta A^j_{(0)} + \ell^2 F^{i}_{(0)u} \delta A_u^{(0)} \, ,\\
        \widetilde{\mathscr{L}}_{(4)} &= - \frac{1}{2} \left( F_{ij}^{(0)} F^{ij}_{(0)} + \ell^2 F^{i}_{(0)u} F_{iu}^{(0)} \right) ,
    \end{align}
\end{subequations}
since one can check that
\begin{equation}
    \widetilde{\Theta}^r_{(3)} \approx \delta \widetilde{\mathscr{L}}_{(4)} - \partial_u \widetilde{\Theta}^u_{(4)} - \partial_i \widetilde{\Theta}^i_{(4)} \, .
\end{equation}
Taking the limit $r \to \infty$, we then have 
\begin{equation}
    \Theta^r_{\text{ren}} = \sqrt{-\gamma} \, \widetilde{\Theta}^r_{(4)} \, ,
\end{equation}
where we emphasize that we recover the result \eqref{delta Sren D=6 any gauge} upon subtracting the boundary term
\begin{equation}
    \ell^4 \partial_u F_{iu}^{(0)} \delta F^i_{(0)u} = \frac{\ell^4}{2} \delta \left( \partial_u F_{iu}^{(0)} F^i_{(0)u} \right) .
\end{equation}
Proceeding in this way, we are led to the same expression for the charge as in \eqref{QBondi6}.

\subsubsection{$D=5$}

In this case, the asymptotic solution space of the Maxwell field is
\begin{equation}
        \mathcal{A}_i = A_i^{(0)} + \tfrac{1}{r} A_i^{(1)} + \mathcal{O}(\tfrac{1}{r^2}) \, , \qquad \mathcal{A}_u = A_u^{(0)} + \mathcal{O}(\tfrac{1}{r^2}) \, .
\end{equation}
The zeroth orders are arbitrary while the first subleading term is given by
\begin{equation}
    A_{i}^{(1)} = \ell^2 \left( \partial_i A_u^{(0)} - \partial_u A_i^{(0)} \right) .
\end{equation}
This implies the following expansion
\begin{equation}
     \Theta^r = \log r \, \sqrt{-\gamma} \, \widetilde{\theta}^r_{(3)} + \sqrt{-\gamma} \, \widetilde{\Theta}^r_{(3)} + \mathcal{O}(\tfrac{1}{r}) \, ,
\end{equation}
that we have to renormalize. Indeed, in the limit $r \to \infty$, the logarithmic term diverges
\begin{equation}
    \widetilde{\theta}^r_{(3)} = \partial^i F_{ij}^{(0)} \delta A^j_{(0)} + \ell^2 F_{ui}^{(0)} \delta F^i_{(0)u} \, .
\end{equation}
Using \eqref{Spin-1 Bondi Presymp Radial Eq order}, we can cancel the latter via ambiguities
\begin{equation}
    \Theta^r_{\text{ct}} = - \log r \, \sqrt{-\gamma} \left( \delta \widetilde{\mathscr{L}}_{(4)} - \partial_u \widetilde{\Theta}^u_{(4)} - \partial_i \widetilde{\Theta}^i_{(4)} \right) ,
\end{equation}
where the boundary and corner terms are respectively
\begin{subequations}
    \begin{align}
        \widetilde{\mathscr{L}}_{(4)} &= - \frac{1}{2} \left( F_{ij}^{(0)} F^{ij}_{(0)} + \ell^2 \gamma^{ij} F^{i}_{(0)u} F_{iu}^{(0)} \right) , \\
        \widetilde{\Theta}^u_{(4)} &= \ell^2 F_{iu}^{(0)} \delta A^i_{(0)} \, ,\\
        \widetilde{\Theta}^i_{(4)} &= - F^{i}_{(0)j} \delta A^j_{(0)} + \ell^2 F^{i}_{(0)u} \delta A_u^{(0)} \, .
    \end{align}
\end{subequations}
Therefore the renormalized presymplectic potential is given by
\begin{equation}
    \Theta^r_{\text{ren}} = \lim_{r\to\infty} \left( \Theta^r + \Theta^r_{\text{ct}} \right) = \sqrt{-\gamma} \, \widetilde{\Theta}^r_{(3)} \, ,
\end{equation}
where
\begin{equation}
    \widetilde{\Theta}^r_{(3)} = \frac{1}{\ell^2} F_{ir}^{(3)} \delta A^i_{(0)} + F_{ru}^{(3)} \delta A_u^{(0)} + \ell^2 \partial_u F_{iu}^{(0)} \delta A^i_{(0)} \, .
\end{equation}
It coincides, without extra term, with the result \eqref{delta Sren D=5 any gauge} obtained by renormalizing the variational principle.
Eventually, this leads to the charge in \eqref{QBondi5}.

\section{Further observations} \label{sec:further}

\subsection{Dictionary between Bondi and Poincar\'e}
We would like to compare our results obtained in the two different coordinate systems presented in the previous sections. To this end, let us write down the associated change of coordinates. We introduce a splitting of the index $I$, which labels the $D-1$ spatial directions $1,\ldots,D-1$ by breaking it down as $I =(I', D-1)$ with $I'=1,\ldots,D-2$. Similarly, we break down the unit vector $\hat X^I(x^i)$ parametrized by the $D-2$ angles $x^i$, with $i=1,\ldots,D-2$, by isolating the angle with respect to the $(D-1)$th direction, so that $x^i=(x^{i'},x^{i=D-2}=\theta)$ with $i'=1,\ldots,D-3$ and
\begin{equation}
    \hat X^I(x^i) = \left(\begin{matrix} \sin\theta\,\hat X^{I'}(x^{i'}) \\ \cos\theta \end{matrix}\right)\,,\qquad
    \hat X^{I'} \hat X^{I'} = 1\,.
\end{equation}
Then we can express the Poincar\'e coordinates in terms of the Bondi coordinates as follows
\begin{align}
    z^{-1} &= \cos\frac{u}{\ell}-\frac{r}{\ell}\left( \cos\theta  + \sin\frac{u}{\ell}\right),\\
    x^0 &= \ell \, \frac{\cos\frac{u}{\ell}+\frac{\ell}{r}\sin\frac{u}{\ell}}{\frac{\ell}{r}\cos\frac{u}{\ell}-\left( \cos\theta  + \sin\frac{u}{\ell}\right)}\,,\\
    x^{I'} &= \ell \, \frac{\hat X^{I'}(x^{i'}) \sin\theta}{\frac{\ell}{r}\cos\frac{u}{\ell}-\left( \cos\theta  + \sin\frac{u}{\ell}\right)}\,.
\end{align}
For $\ell\ll r$, one finds
\begin{equation} \label{roughscalings}
    z \sim \frac{\ell}{r(-\sin\frac{u}{\ell}-\cos\theta)}\,,\qquad
    x^0 \sim \frac{\ell \cos\frac{u}{\ell}}{(-\sin\frac{u}{\ell}-\cos\theta)}\,,\qquad
    x^{I'} \sim \frac{\ell \sin\theta \hat X^{I'}(x^{i'})}{(-\sin\frac{u}{\ell}-\cos\theta)}
\end{equation}
so that in particular $z\propto 1/r$, provided
\begin{equation}
    \cos\theta<-\sin\frac{u}{\ell}\,.
\end{equation}
This relation highlights rather clearly that, depending on the retarded time at which one approaches the boundary, only a portion of the sphere at infinity is covered by the Poincar\'e coordinates, except when $u=-\pi\ell/2$ (i.e.~$X^0=0$ for large $r$) in which case the whole sphere is covered except for the North pole (see Fig.~\ref{fig:AdS3Penrose}).
Using the change of coordinates given above we can calculate the component transformation for the field strength for large~$r$. In particular, we find
\begin{equation} \label{urzaMapping}
    \mathcal F_{ru}
    \sim
    \ell\,\frac{\mathcal F_{z0} \left( 1 + \cos\theta \sin\frac{u}{\ell}\right) + \mathcal F_{zI'} \hat X^{I'}  \sin\theta\,\cos\frac{u}{\ell}}{r^2(\sin\frac{u}{\ell}+\cos\theta)^3}\,.
\end{equation}
We also note that, for field configurations such that, for small $z$ in Poincar\'e coordinates 
\begin{equation} \label{CoulombicLeading}
    \mathcal F_{z0} \sim z^{D-4} F_{z0}^{(D-4)}\,,
\end{equation}
which as we have seen play the role of the ``VEV'' or ``Coulombic'' branch in the analysis of the equations of motion, then, at $u=-\pi \ell/2$,
\begin{equation}\label{FFcorrespondence}
\mathcal F_{ru} \sim \frac{F_{ru}^{(D-2)}}{r^{D-2}}\,,\qquad
F^{(D-2)}_{ru} = - \frac{\ell^{D-3} F_{z0}^{(D-4)}}{(1-\cos\theta)^{D-2}}\,.
\end{equation}
Here the minus sign and the offset $D-4\to D-2$ reflect the fact that $z$ roughly scales as the inverse of $r$ (see Eq.~\eqref{roughscalings}).
Moreover, by using \eqref{FFcorrespondence} and noting that
\begin{equation}
    \text{d}^{D-2}x = \frac{\ell^{D-2} (\sin\theta)^{D-3}}{(1-\cos\theta)^{D-2}}\sqrt{\gamma'}\,\text{d}\theta\,\text{d}^{D-3}x\,,\qquad
    \text{d}\Omega_{D-2}(\hat X) = (\sin\theta)^{D-3}\sqrt{\gamma'}\,\text{d}\theta\,\text{d}^{D-3}x\,,
\end{equation}
with $\gamma'= \operatorname{det}(\gamma_{i'j'})=\operatorname{det}(\partial_{i'}\hat X^{I'} \partial_{j'}\hat X^{I'})$,
one can explicitly check that
\begin{equation} \label{equalcharges}
    \frac{1}{\ell} \int_{\substack{x^0=0\\z=0}} \lambda\,  F_{z0}^{(D-4)} \,\text{d}^{D-2}x 
    =
    \int_{\substack{u=-\pi\ell/2\\r=\infty}}  \lambda \, F^{(D-2)}_{ur}\, \text{d}\Omega_{D-2}(\hat X) .
\end{equation}
This shows that the charges constructed in the two setups are indeed identical, as expected. 

Let us now briefly comment on the role of the independent free functions appearing in the solutions of the equations of motion. The relation between the ``VEV'' or ``Coulombic'' branches in the two coordinate systems is discussed above. The information about the ``source'' or radiation branch, in Poincar\'e coordinates is contained in the leading, $\mathcal O(z^0)$ component of $\mathcal F_{ab}$,
\begin{equation}
    \mathcal F_{ab} \sim F_{ab}^{(0)}\,.
\end{equation}
This translates into Bondi components via
\begin{equation} \label{ijabMappping}
    \begin{split}
        \mathcal F_{uj'} &\sim \ell \frac{\partial \hat X^{J'}}{\partial x^{j'}}
        \frac{\sin\theta\left[-\mathcal F_{0J'} 
        \left( 1 + \cos\theta \sin\frac{u}{\ell}\right) 
        + \mathcal F_{J'I'} \hat X^{I'}  \sin\theta\,\cos\frac{u}{\ell}\right]}{(\sin\frac{u}{\ell}+\cos\theta)^3}\,,
        \\
        \mathcal F_{u\theta} & \sim -\ell\, \frac{\mathcal F_{0I'}\hat X^{I'}}{(\sin\frac{u}{\ell}+\cos\theta)^2}\,,\\
        \mathcal F_{i'j'} &\sim \ell^2\, \frac{\partial \hat X^{I'}}{\partial x^{i'}} \mathcal F_{I'J'}  \frac{\partial \hat X^{J'}}{\partial x^{j'}} \frac{(\sin\theta)^2}{(\sin\frac{u}{\ell}+\cos\theta)^2}\,,
        \\
        \mathcal F_{i'\theta} &\sim \ell^2\, \frac{\partial \hat X^{I'}}{\partial x^{i'}} \frac{\sin\theta
        \left[
        \mathcal F_{I'J'} \hat X^{J'} 
        \left( 1 + \cos\theta \sin\frac{u}{\ell}\right) 
        - \mathcal F_{0I'} \sin\theta\,\cos\frac{u}{\ell}
        \right]
        }{(\sin\frac{u}{\ell}+\cos\theta)^3} \, ,
    \end{split}
\end{equation}
so inducing $\mathcal O(r^0)$ terms in $\mathcal F_{ui}$ and $\mathcal F_{ij}$, which thus encode the dependence on the radiation data. Note that Eq.~\eqref{ijabMappping} does not involve any additional factor of $1/r$ to leading order, in contrast with \eqref{urzaMapping}. 

\subsection{Flat limit}
In order to take the flat-space limit $\ell\to\infty$, we need to fix a choice of coordinate system, and the factor of $\frac{1}{\ell}$ on the left-hand side of \eqref{equalcharges} suggests that the flat limit be better defined in Bondi coordinates than in Poincar\'e ones.

Actually, from \eqref{charge5B} for $D=5$ and \eqref{charge6B} for $D=6$, we see that even in these more convenient coordinates the symplectic structure itself still contains potentially dangerous terms that scale as $\ell^2$. However, such offending terms actually are total derivatives on the sphere and vanish identically. We can then safely take the limit $\ell\to\infty$ and find the standard expression 
\begin{equation}\label{rightQ}
    Q^{(\lambda)} = - \int \text{d}^{D-2}x \left(\lambda \, F_{ur}^{(D-2)}\right) .
\end{equation}
Whereas we have explicitly performed the derivation in Bondi coordinates only for $D=5,6$, but our generic-$D$ results of section~\ref{sec:Spin1Poincare}, in particular the charge \eqref{ChargeExplEvenP}, combined with \eqref{equalcharges} allow us to conclude that the expression \eqref{rightQ} applies in any dimensions, for both AdS and flat spacetime. 

\acknowledgments

We would like to thank Luca Ciambelli for useful discussions.
Let us also thank the anonymous referee for comments that helped us improve the manuscript.
We are grateful to Raffaele Marotta for correspondence and for useful observations that helped us correct several misprints.
The research of AC and AD is partially supported by the Fonds de la Recherche Scientifique - FNRS under Grants No.\ FC.41161, F.4503.20 and T.0022.19.
The research of CH is supported by the
Knut and Alice Wallenberg Foundation under grant KAW 2018.0116. Nordita is partially
supported by Nordforsk. This work was funded by UK Research and Innovation (UKRI) under the UK government’s Horizon Europe funding guarantee [grant number EP/X037312/1].

\appendix

\section{Geometry of AdS}\label{sec:AdS}

In this appendix, we collect a few facts about the geometry of AdS spacetime and 
detail the coordinate systems that are used in the main body of the text.
A standard presentation of $D$-dimensional AdS spacetime can be obtained starting from $(D+1)$-dimensional flat space with line element
\begin{equation}\label{}
	\text{d}s^2 = \text{d}X^M \eta_{MN}\, \text{d}X^N\,,
 \qquad
    \eta_{MN}=\mathrm{diag}(-1,1,\ldots,1,- 1)\,,
\end{equation}
where $N,M=0,1,\ldots,D-1,D$, and imposing the following constraint ($\ell$ being the AdS radius)
\begin{equation}\label{constraint}
	X^M \eta_{MN} X^N = -\ell^2\,.
\end{equation}

\subsection{Global coordinates and Penrose diagram}
\begin{figure}
	\centering
	\begin{tikzpicture}[scale=1.1]
 \draw[help lines, color=gray!40,thick,->] (-4,0) -- (4,0);
	\node[right] at (4,0) {$X^0$};
	\draw[help lines, color=gray!40,thick,->] (0,-3.5) -- (0,3.5);
	\node[right] at (0,3.5) {$X^1$};
 \draw[help lines, color=gray!40,thick,->] (3,.5) -- (-3,-.5);
	\node[left] at (-3,-.5) {$X^2$};
  \draw (-3,2) .. controls (-.65,.5) and (-.65,-.5) .. (-3,-2);
  \draw (3,2) .. controls (.65,.5) and (.65,-.5) .. (3,-2);
  \draw [gray!50] (-3,2) .. controls (-3,1.75) and (3,1.75) .. (3,2);
  \draw [gray!50](-3,2) .. controls (-3,2.25) and (3,2.25) .. (3,2);
  \draw[red,dashed] (-1.24,0) .. controls (-1.24,.2) and (1.24,.2) .. (1.24,0);
  \draw[red] (-1.24,0) .. controls (-1.24,-.2) and (1.24,-.2) .. (1.24,0);
  \draw[green!70!black] (-2,1.85) .. controls (-.35,.1) and (-.35,-.3) .. (-2,-2.15);
  \draw[green!70!black,dashed] (2,-1.85) .. controls (.35,-.1) and (.35,.3) .. (2,2.15);
  \draw[thick,violet] (-1.12,1.82) -- (1.7,-1);
  \draw[thick,densely dotted,violet] (1.7,-1) -- (2.6,-1.9);
  \draw[thick,violet] (-2.59,1.89) -- (-1.8,1.1);
  \draw[thick,densely dotted,violet] (-1.8,1.1) -- (1.1,-1.8);
  \draw[gray!50,dashed] (-3,-2) .. controls (-3,-1.75) and (3,-1.75) .. (3,-2);
  \draw [gray!50] (-3,-2) .. controls (-3,-2.25) and (3,-2.25) .. (3,-2);
  \draw [red,->] (-.7,-.25) .. controls (-.6,-.28) and (-.3,-.29) .. (-.2,-.29);
  \node[below] at (-.2,-.29) {{\color{red}T}};
  \draw [green!70!black,->] (-.7,.25) .. controls (-.71,.26) and (-.77,.49) .. (-1,.83);
  \node[right] at (-1,.83) {{\color{green!70!black}R}};
	\end{tikzpicture}
	\caption{\label{fig:AdS2Global} A representation of AdS$_2$. The red line is the $R=0$ submanifold, the solid green line is the one at $T=0$ and the dashed green line is the one at $T=\pi\ell$. The  purple lines mark the intersection with the plane $X^1 = X^2$: the portion of the spacetime  below this plane is covered by the Poincar\'e coordinates, see Eq.~\eqref{PoincareRange}.}
\end{figure}
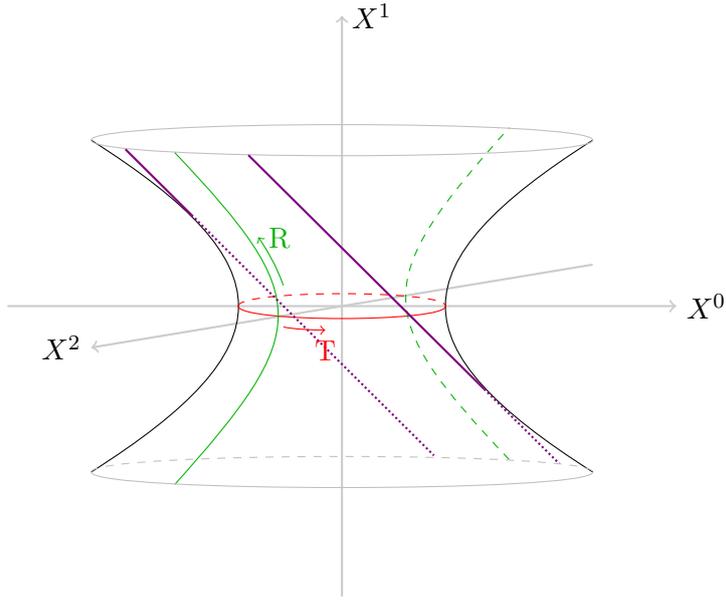

Introducing a time coordinate $T$, a radial coordinate $R$ and $D-2$ angular coordinates $x^i$ by 
\begin{equation}
X^0 = \ell \cosh\frac{R}{\ell} \,\sin\frac{T}{\ell}\,,\qquad
X^I = \ell \,\hat X^I(x^i) \,\sinh\frac{R}{\ell}\,,\qquad
X^D = \ell \cosh \frac{R}{\ell}\, \cos\frac{T}{\ell}\,,
\end{equation}
with $\hat X^I$ a  Euclidean unit vector,
$\hat X^I \hat X^I = 0$,
the metric takes the form 
\begin{equation}
    \text{d}s^2 = - (\cosh\tfrac{R}{\ell})^2 \text{d}T^2 + \text{d}R^2 + \ell^2 (\sinh\tfrac{R}{\ell})^2 \text{d}\Omega^2\,,\quad
  \text{d}\Omega^2=\text{d}x^i\gamma_{ij}\, \text{d}x^j\,,
 \quad 
 \gamma_{ij} = \frac{\partial \hat X^I}{\partial x^i}\frac{\partial \hat X^I}{\partial x^j}\,.
\end{equation}
The main advantage of the set of coordinates $(T,R,x^i)$ is that they cover the whole AdS$_D$ spacetime, see Fig.~\ref{fig:AdS2Global} for a schematic representation in the case $D=2$.

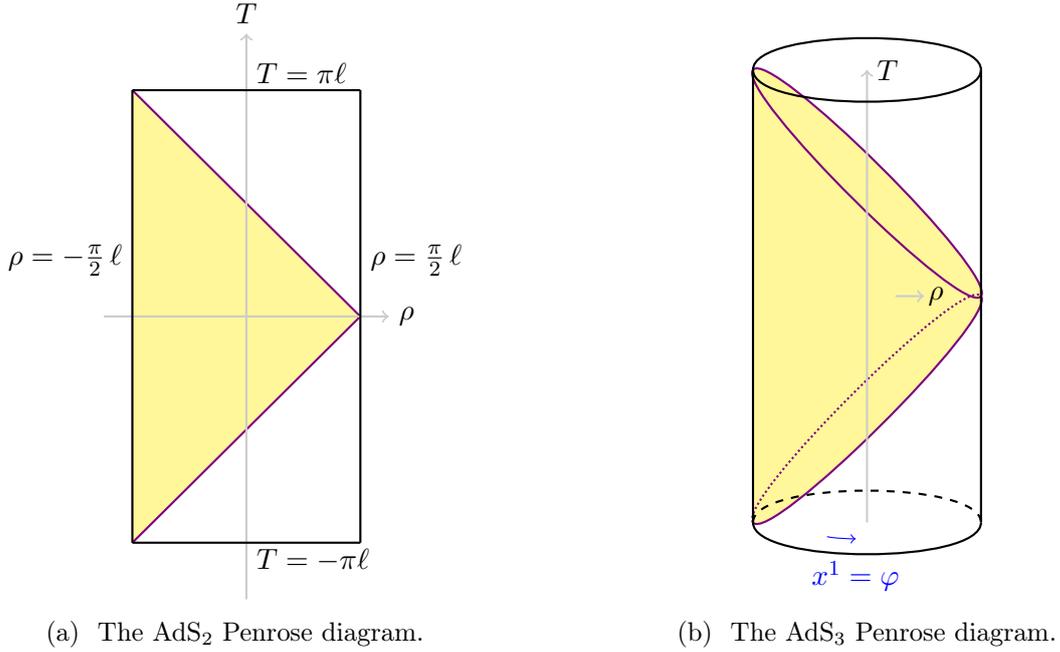
\begin{figure}
	\centering
	\begin{subfigure}{.45\textwidth}
		\centering
		\begin{tikzpicture}[scale=.75]
    \fill[fill=white!50!yellow] (-2,4) -- (-2,-4) -- (2,0) -- cycle;
    \draw [thick,violet] (-2,4) -- (2,0);
    \draw [thick,violet] (-2,-4) -- (2,0); 
    \draw[help lines, color=gray!40,thick,->] (-2.5,0) -- (2.5,0);
	\node[right] at (2.5,0) {$\rho$};
	\draw[help lines, color=gray!40,thick,->] (0,-5) -- (0,5);
	\node[above] at (0,5) {$T$};
		\draw[thick] (-2,-4)--(-2,4);
    \node[left] at (-2,1) {$\rho = -\frac{\pi}{2}\,\ell$};
        \draw[thick] (2,4)--(2,-4);
    \node[right] at (2,1) {$\rho = \frac{\pi}{2}\,\ell$};
        \draw[thick] (-2,-4)--(2,-4);
    \node[right] at (0,4.3) {$T = \pi \ell$};
        \draw[thick] (-2,4)--(2,4);
    \node[right] at (0,-4.3) {$T = -\pi \ell$};		
    \end{tikzpicture}
	\caption{\label{fig:AdS2Penrose} The AdS$_2$ Penrose diagram.}
	\end{subfigure}
\hfill
\begin{subfigure}{.45\textwidth}
	\centering
	\begin{tikzpicture}[scale=.75]
 \fill[fill=white!50!yellow] (-2,4) -- (-2,-4) -- (2,0) -- cycle;
 \draw[thick,violet, fill=white!50!yellow] (-2,4) .. controls (-1.7,4.4) and (2.3,.4) .. (2,0);
 \draw[thick,violet, fill=white!50!yellow] (-2,-4) .. controls (-1.7,-4.4) and (2.3,-.4) .. (2,0);
 \draw[thick,violet] (-2,4) .. controls (-2.3,3.6) and (1.7,-.4) .. (2,0);
 \draw[thick,densely dotted,violet] (-2,-4) .. controls (-2.3,-3.6) and (1.7,.4) .. (2,0);
    \draw[help lines, color=gray!40,thick,->] (.5,0) -- (1,0);
	\node[right] at (.9,0) {$\rho$};
    \draw[help lines, color=gray!40,thick,->] (0,-4) -- (0,4);
	\node[right] at (0,4) {$T$};
		\draw[thick] (-2,-4)--(-2,4);
        \draw[thick] (2,4)--(2,-4);
        \draw[thick] (-2,4) .. controls (-2,3.25) and (2,3.25) .. (2,4);
        \draw[thick] (-2,4) .. controls (-2,4.75) and (2,4.75) .. (2,4);
        \draw[thick, dashed] (-2,-4) .. controls (-2,-3.25) and (2,-3.25) .. (2,-4);
        \draw[thick] (-2,-4) .. controls (-2,-4.75) and (2,-4.75) .. (2,-4);
        \draw [blue,->] (-.7,-4.25) .. controls (-.6,-4.28) and (-.3,-4.29) .. (-.2,-4.29);
        \node[below] at (-.2,-4.5) {{\color{blue}$x^1=\varphi$}};
	\end{tikzpicture}
	\caption{\label{fig:AdS3Penrose} The AdS$_3$ Penrose diagram.}
\end{subfigure}
	\caption{\label{fig:AdS23Penrose} Penrose diagrams for AdS$_2$ and AdS$_3$. The highlighted region is the portion of spacetime covered by the Poincar\'e coordinate patch (see Eq.~\eqref{PoincareRange}).
 The boundary of this region has two components (see below Eq.~\eqref{PoincareRange}): $z=0$, which lies on the actual boundary of AdS$_D$, and $z=\infty$, which lies in the interior of AdS$_D$.}
\end{figure}

Introducing the compactified radial variable
\begin{equation}
    \rho = \ell \arctan\left( \sinh\frac{R}{\ell}\right)\,,
\end{equation}
the metric becomes
\begin{align}
    \text{d}s^2 = \frac{1}{(\cos\tfrac{\rho}{\ell})^2}\left[
    -\text{d}T^2 + \text{d}\rho^2 + \ell (\sin\tfrac{\rho}{\ell})^2 \text{d}\Omega^2
    \right]
    \end{align}
from which it becomes manifest that the conformal boundary is the surface of a cylinder located at $\rho = \pi\ell/2$ as depicted in Fig.~\ref{fig:AdS23Penrose}. 
Although the original time coordinate $T$ is periodic, $T\sim T+2\pi \ell$, as is clear from Fig.~\ref{fig:AdS2Global}, in order to avoid closed time-like loops it is standard prolong this cylinder to infinity and to adopt a decompactified coordinate $-\infty < T < +\infty$.

\subsection{Poincar\'e coordinates} \label{app:Poincare}

One can alternatively solve the defining constraint \eqref{constraint} by introducing Poincar\'e coordinates $(z,x^a)$, 
with $a=0,1,\ldots,D-1$,
as follows
\begin{equation}\label{PoincareCoordinates}
X^a = \frac{x^a}{z}\,,\qquad
X^{D-1} = z\,\frac{\ell}{2} \left(1+\frac{ x^2}{\ell^2 z^2}-\frac{1}{z^2}\right),\qquad
X^D = z\,\frac{\ell}{2}\left(1+\frac{x^2}{\ell^2z^2}+\frac{1}{z^2}\right),
\end{equation}
where $x^2=x^a \eta_{ab}\, x^b$ and  $\eta_{ab}=\mathrm{diag}(-1,1,\ldots,1)$.
The AdS metric then takes the form
\begin{equation}\label{AdSmetricinPoincare}
	\text{d}s^2 = \frac{\ell^2 \text{d}z^2 + \text{d}x^a \eta_{ab} \text{d}x^b}{z^2}\,.
\end{equation}
Since
\begin{equation}\label{PoincareRange}
    X^{D}-X^{D-1} = \frac{\ell}{z}\,,\qquad z>0\,,
\end{equation}
the Poincar\'e coordinates only cover the portion of AdS space lying in the half-space $X^D>X^{D-1}$ (see Figs.~\ref{fig:AdS2Global}, \ref{fig:AdS23Penrose}).
As $z\to\infty$, one approaches the hyperplane $X^D = X^{D-1}$ slicing the AdS space in two, while, as $z\to0$, one approaches the actual boundary of AdS. The metric determinant and the inverse metric take the following forms
\begin{equation}
\sqrt{-g} = \frac{\ell}{z^D}\,,\qquad
g^{\mu\nu} \partial_\mu \partial_\nu = z^2 \left( \ell^{-2} \partial_z^2 + \eta^{ab} \partial_a \partial_b \right),
\end{equation}
while the non-zero Christoffel symbols are
\begin{equation}
\Gamma^z_{zz} = - \frac{1}{z} \, , \qquad \Gamma^z_{ab} = \frac{1}{z \ell^2}\, \eta_{ab} \, , \qquad \Gamma^a_{b z} = - \frac{1}{z}\, \delta^a_b \, .
\end{equation}

\subsection{Bondi coordinates}
\label{app:Bondi}

Starting again from the embedding space, one can introduce polar coordinates for the spatial directions
$I=1,2\ldots,D-1$ according to 
$X^I = r \hat X^I(x^i)$, with $\hat X^I \hat X^I=1$,
and $i=1,2,\ldots,D-2$ labeling the angular variables. One can then solve the constraint \eqref{constraint} by letting
\begin{equation}\label{}
	X^0=\ell \sqrt{1+\left(\frac{r}{\ell}\right)^2}\sin\left(\frac{u}{\ell}+\arctan\frac{r}{\ell}\right),\qquad
	X^D=\ell \sqrt{1+\left(\frac{r}{\ell}\right)^2}\cos\left(\frac{u}{\ell}+\arctan\frac{r}{\ell}\right).
\end{equation}
In this way, $(u,r,x^i)$ define Bondi coordinates on AdS$_D$ and the metric takes the form \cite{Poole:2018koa}
\begin{equation}\label{}
	\text{d}s^2 = -\left(1+\frac{r^2}{\ell^2}\right)\text{d}u^2-2\text{d}u \text{d}r + r^2\text{d}\Omega^2\,,\qquad  \text{d}\Omega^2=\text{d}x^i\gamma_{ij}\, \text{d}x^j\,,
 \qquad 
 \gamma_{ij} = \frac{\partial \hat X^I}{\partial x^i}\frac{\partial \hat X^I}{\partial x^j}\,.
\end{equation}
The determinant is given by $\sqrt{-g} = r^{D-2} \sqrt{-\gamma}$ and the inverse metric by
\begin{equation}
    g^{\mu\nu} \partial_\mu \partial_\nu = - 2 \partial_u \partial_r + \left(1 +  {\frac{r^2}{\ell^2}}\right) \partial_r^2 + r^{-2} \gamma^{ij} \partial_i \partial_j \, .
\end{equation}
The non-zero Christoffel symbols are
\begin{equation}
\begin{aligned}
&\Gamma^i_{rj} = \frac{1}{r}\, \delta^i_j \, , && {\Gamma^u_{uu} = - \frac{r}{\ell^2}} \, , &&&&\Gamma^u_{ij} = r\, \gamma_{ij} \, ,\\
& {\Gamma^r_{ru} = \frac{r}{\ell^2}} \, , && {\Gamma^r_{uu} = \frac{r}{\ell^2} \left(1 + \frac{r^2}{\ell^2}\right)} , &&&&\Gamma^r_{ij} = - r \left(1 +  {\frac{r^2}{\ell^2}}\right) \gamma_{ij} \, ,\\
&\Gamma^i_{jk} = \tfrac{1}{2}\, \gamma^{il} \left( \partial_j \gamma_{kl} + \partial_k \gamma_{jl} - \partial_l \gamma_{jk} \right) .
\end{aligned}
\end{equation}
%

\section{Covariant phase space formalism}\label{app:CPS}

In this appendix, we review the covariant phase space (CPS) formalism in order to introduce the notations used in the main body of the text. This formulation was introduced in \cite{GAWEDZKI1972307,kijowski1973finite,kijowski1976canonical} and refined in \cite{Lee:1990nz,Wald:1993nt,Wald:1999wa,Barnich:2001jy}\footnote{See, e.g., \cite{Compere:2018aar,Ruzziconi:2019pzd,Ciambelli:2022vot} for pedagogical reviews.}.

The differentiable manifold that we considered is the Anti de Sitter spacetime $\mathcal{M} = \text{AdS}_D$ of dimension $D$. On the latter, the forms induce the de Rham cohomology, with exterior derivative $\text{d}$ (such that $\text{d}^2 = 0$) and interior product $i$. The Lie derivative along a diffeomorphism $\xi \in T\mathcal{M}$ is given by
\begin{equation}
    \mathcal{L}_\xi = \text{d} i_\xi + i_\xi \text{d} \, .
\end{equation}
The idea of the CPS formalism is to put together spacetime and phase space calculi. The space of all possible field configurations is also a differential manifold $\Gamma$, with $\delta$ (such that $\delta^2 = 0$) and $I$ respectively the exterior derivative and the interior product on the latter. The formula
\begin{equation}
    \mathfrak{L}_V = \delta I_V + I_V \delta
\end{equation}
computes their effects when applied in different orders along $V \in T \Gamma$. A theory is specified by an action
\begin{equation}
    S = \int_{\mathcal{M}} L \, .
\end{equation}
where $L = \mathscr{L} \, \text{d}^Dx $ is the Lagrangian form. An arbitrary field variation yields 
\begin{equation} \label{local presymp pot def}
    \delta L = (\text{eom}) \delta \Psi + \text{d} \Theta \, ,
\end{equation}
where $(\text{eom})$ denotes the equations of motion, $\Psi \in \Gamma$ and $\Theta = \Theta^\mu (\text{d}^{n-1}x)_\mu$ is the local presymplectic potential form. One defines the local presymplectic two-form
\begin{equation} \label{local presymp form def}
    \omega = \delta \Theta \, 
\end{equation}
in terms of which one computes the presymplectic two-form
\begin{equation} \label{presymp form def}
    \Omega = \int_\Sigma \omega \, ,
\end{equation}
as the integral over an (arbitrary) Cauchy surface $\Sigma \subset \mathcal{M}$. The presymplectic potential admits two types of modifications ---~often referred to as ambiguities in the literature although in general they lead to inequivalent presymplectic potentials~---
which do not alter the equations of motion:
\begin{equation} \label{ambiguities def}
    \Theta \to \Theta + \delta B - \text{d} C \, .
\end{equation}
The first one corresponds to the addition of a boundary term to the Lagrangian, $L \to L + \text{d}B$, and does not contribute to the presymplectic form $\omega$ since $\delta^2=0$. The second one is related to the fact that $\Theta$ appears as a boundary term in $\delta L$. The latter has an impact on $\omega$,
\begin{equation}
    \omega \to \omega - \text{d} \delta C =: \omega + \text{d} \omega_C \, ,
\end{equation}
while not affecting $\Omega$. This is in line with the study of the so-called corner terms \cite{Donnelly:2016auv,Speranza:2017gxd,Geiller:2017whh,Geiller:2017xad,Freidel:2020xyx,Freidel:2020svx,Freidel:2020ayo,Donnelly:2020xgu,Ciambelli:2021vnn,Freidel:2021cjp,Ciambelli:2021nmv,Ciambelli:2022cfr}.

A vector $V \in T \Gamma$ is called symplectomorphism or Hamiltonian vector field if $\mathfrak{L}_V \omega = 0$. Since $  \mathfrak{L}_V \omega = \delta I_V \omega$, given a Hamiltonian vector field one can define a current $J_V$ and the associated global functional $H_V$ as
\begin{equation}
    I_V \omega = - \delta J_V \, , \qquad H_V = \int_\Sigma J_V \, .
\end{equation}
If this vector corresponds to a spacetime symmetry $\xi$, we denote it $V = V_\xi$ and we have that $\mathfrak{L}_{V_\xi} = \mathcal{L}_\xi$. This implies that we can write the current as follows,\footnote{In this construction, for simplicity, we assume $\text{d} i_\xi \Theta$ pull-backs to zero at the boundary.}
\begin{equation}
\quad J_{V_\xi} \approx I_{V_\xi} \Theta - i_\xi L \, ,
\end{equation}
modulo $\delta$-exact terms. This current is called the local weakly-vanishing Noether current. If, in addition, the vector $V_\xi$ is a gauge symmetry, we have also that $\mathfrak{L}_V S = \mathcal{L}_\xi S = 0$, where
\begin{equation}
    \mathfrak{L}_{V_\xi} S \approx \int_{\partial \mathcal{M}} I_{V_\xi} \Theta \, , \qquad \mathcal{L}_\xi S = \int_{\partial \mathcal{M}} i_\xi L\, .
\end{equation}
Equating the two we get the fundamental theorem of CPS,
\begin{equation} \label{Noether 2nd}
    I_{V_\xi} \Theta \approx i_\xi L \quad \Rightarrow \quad J_{V_\xi} \approx \text{d} Q_\xi \, ,
\end{equation}
stating that for gauge symmetries the Noether charge is a corner term:
\begin{equation}
    H_\xi = \int_\Sigma J_{V_\xi} \approx \int_{S = \partial \Sigma} Q_\xi \, .
\end{equation}
Moreover, one can  add to $Q_\xi$ the divergence of a ($D-3$)--form without modifying $H_\xi$, due to Stokes' theorem.


\providecommand{\href}[2]{#2}
\end{document}